\documentclass[twocolumn]{aastex63}

\definecolor{DarkOrange}{RGB}{204, 85, 0}
\definecolor{LincolnGreen}{RGB}{17, 102, 0}
\definecolor{Rust}{HTML}{9B4F0F}
\definecolor{DarkCyan}{HTML}{008B8B}
\definecolor{MediumAquaMarine}{HTML}{66CDAA}

\usepackage{lineno}
\usepackage{pifont}% http://ctan.org/pkg/pifont

\usepackage{multirow, amsmath}

\received{\today}
\revised{}
\accepted{}
\submitjournal{PASP}

\shorttitle{PS1 Point Source Catalog II}
\shortauthors{Miller \& Hall}
\watermark{DRAFT}
\graphicspath{{./}{figures/}}
\begin{document}

\title{A Morphological Classification Model to Identify Unresolved PanSTARRS1 Sources II: Update to the PS1 Point Source Catalog}

\author[0000-0001-9515-478X]{A.~A.~Miller}
\affiliation{Center for Interdisciplinary Exploration and Research in 
             Astrophysics (CIERA) and Department of Physics and Astronomy, 
             Northwestern University, 
             1800 Sherman Road, Evanston, IL 60201, USA}
\affiliation{The Adler Planetarium, Chicago, IL 60605, USA}
\email{amiller@northwestern.edu}

\author[0000-0002-9364-5419]{X.~Hall}
\affiliation{Cahill Center for Astrophysics,
             California Institute of Technology,
             1200 E.~California Boulevard, Pasadena, CA 91125, USA}
\affiliation{Center for Interdisciplinary Exploration and Research in 
             Astrophysics (CIERA) and Department of Physics and Astronomy, 
             Northwestern University, 
             1800 Sherman Road, Evanston, IL 60201, USA}

%% Note that the \and command from previous versions of AASTeX is now
%% depreciated in this version as it is no longer necessary. AASTeX 
%% automatically takes care of all commas and "and"s between authors names.

%% AASTeX 6.3 has the new \collaboration and \nocollaboration commands to
%% provide the collaboration status of a group of authors. These commands 
%% can be used either before or after the list of corresponding authors. The
%% argument for \collaboration is the collaboration identifier. Authors are
%% encouraged to surround collaboration identifiers with ()s. The 
%% \nocollaboration command takes no argument and exists to indicate that
%% the nearby authors are not part of surrounding collaborations.

%% Mark off the abstract in the ``abstract'' environment. 
\begin{abstract}

We present an update to the PanSTARRS-1 Point Source Catalog (PS1 PSC), which
provides morphological classifications of PS1 sources. The original PS1 PSC
adopted stringent detection criteria that excluded hundreds of millions of PS1
sources from the PSC. Here, we adapt the supervised machine learning methods
used to create the PS1 PSC and apply them to different photometric
measurements that are more widely available, allowing us to add $\sim$144
million new classifications while expanding the the total number of sources in
PS1 PSC by $\sim$10\%. We find that the new methodology, which utilizes PS1
forced photometry, performs $\sim$6--8\% worse than the original method. This
slight degradation in performance is offset by the overall increase in the
size of the catalog. The PS1 PSC is used by time-domain surveys to filter
transient alert streams by removing candidates coincident with
point sources that are likely to be Galactic in origin. The addition of
$\sim$144 million new classifications to the PS1 PSC will improve the
efficiency with which transients are discovered.

\end{abstract}

%% Keywords should appear after the \end{abstract} command. 
%% See the online documentation for the full list of available subject
%% keywords and the rules for their use.

%% From the front matter, we move on to the body of the paper.
%% Sections are demarcated by \section and \subsection, respectively.
%% Observe the use of the LaTeX \label
%% command after the \subsection to give a symbolic KEY to the
%% subsection for cross-referencing in a \ref command.
%% You can use LaTeX's \ref and \label commands to keep track of
%% cross-references to sections, equations, tables, and figures.
%% That way, if you change the order of any elements, LaTeX will
%% automatically renumber them.
%%
%% We recommend that authors also use the natbib \citep
%% and \citet commands to identify citations.  The citations are
%% tied to the reference list via symbolic KEYs. The KEY corresponds
%% to the KEY in the \bibitem in the reference list below. 

\section{Introduction} \label{sec:intro}

The proliferation of wide-field time-domain surveys over the past $\sim$decade
has led to the discovery of a bevy of novel extragalactic transients
\citep[e.g.,][]{quimby11,Gezari12,Drout14,Gal-Yam14,Abbott17a,Prentice18,
IceCube-Collaboration18}. While these wide-field surveys have been enabled by
significant advances in detector technology, software has proven equally
important \citep[e.g.,][]{Masci17,Masci19,Smith20,Jones20} as many of these
critical discoveries have been facilitated by the rapid identification and
dissemination of new transient candidates in near real time
\citep[e.g.,][]{Patterson19}.

Reliable catalogs identifying stars and galaxies, or similarly unresolved and
resolved sources, are an essential cog in the machinery necessary to identify
extragalactic transients. On a nightly basis, time-domain surveys are inundated
with transient candidates, the vast majority of which are considered ``bogus''
\citep[e.g.,][]{Bloom12}. Despite sophisticated software capable of whittling
down the number of likely transients by several orders of magnitude
\citep[e.g.,][]{Brink13,Goldstein15,Duev19,Smith20}, the number of candidates
still vastly outpaces the spectroscopic resources necessary to classify
everything that varies \citep[e.g.,][]{Kulkarni20}. The aforementioned
star--galaxy catalogs therefore play an essential role in the search for
transients by removing stellar-like objects that are likely to be Galactic in
origin.

The PanSTARRS-1 Point Source Catalog \citep[PS1 PSC;][]{Tachibana18}, which
provides probabilistic point-source like classifications for $\sim$1.5 billion
sources detected by PanSTARRS-1 \citep[PS1;][]{Chambers16}, was designed
precisely to filter such sources. This catalog has been deployed by the Zwicky
Transient Facility \citep[ZTF;][]{Bellm19} and other surveys
\citep{Smith20,Moller20} to identify likely extragalactic transients. The PS1
PSC has been demonstrated to be an important ingredient in the systematic
search for extragalactic transients \citep[e.g.,][]{Fremling20,De20}.

A downside to the PS1 PSC is that it does not provide classifications for
sources that are not ``detected'' in the PS1 \texttt{StackObjectAttributes}
table \citep[see \S3 in][]{Tachibana18}. Of the $\sim$3 billion unique sources
in the PS1 \texttt{StackObjectAttributes} table, the vast majority of those
missing from the PS1 PSC are either spurious or have an extremely low
signal-to-noise ratio (S/N), such that the methods in \citet{Tachibana18}
would not provide a reliable classification. Additional sources are missing
from the PS1 PSC because there are multiple rows within the PS1
\texttt{StackObjectAttributes} table that have the same \texttt{ObjID} and
$\mathtt{primaryDetection} = 1$. By definition this should not happen, and
therefore these sources were excluded. For PS1 sources that are not in the PS1
PSC, ZTF reports a probability score $=0.5$, i.e., an ambiguous
classification, when cross-matching newly observed variables with the PS1
catalog (see \ref{app:cat_counts} for additional details about which PS1
sources are used by ZTF).

Here, we present an update to the PS1 PSC by classifying $\sim$144 million
sources that were previously ``missing'' from the catalog. These
classifications are made using different photometric measurements from the
ones adopted in \citet{Tachibana18}. While our new method performs slightly
worse than the one in \citet{Tachibana18}, we nevertheless achieve a similar
level of accuracy with the new model. We apply our new model to the $\sim$426
million ``missing'' sources (classifying $\sim$34\% of them), providing a new
and useful supplement to the PS1 PSC.\footnote{During the preparation of this
manuscript \citet{Beck20} published a new machine learning catalog (PS1-STRM)
to classify the $\sim$2.9 billion sources in the PS1 \texttt{ForcedMeanObject}
table. We highlight differences and similarities between the
\citeauthor{Beck20} catalog and this work in \S\ref{sec:discussion}.}

Alongside this paper, we have released our open-source software needed to
recreate the analysis in this study. These are available online at
\url{https://github.com/adamamiller/PS1_star_galaxy}.

\section{ML Model Data}

\subsection{PS1 Data}

PS1 conducted a five filter ($g_\mathrm{PS1}$, $r_\mathrm{PS1}$,
$i_\mathrm{PS1}$, $z_\mathrm{PS1}$, $y_\mathrm{PS1}$) time-domain survey
covering $\sim$3/4 of the sky \citep{Chambers16}. PS1 provides three different
types of photometric measurements: there are mean flux measurements from the
individual PS1 exposures of each field, there are stack flux measurements from
the deeper stack images that co-add individual exposures, and there are
forced-flux measurements that measure the flux in individual exposures at the
location of all sources detected in the stack images. The mean photometry is
limited by the depth of the individual exposures, while the stack photometry
has a difficult to model point spread function (PSF) because images must be
warped before they can be co-added. The forced-flux measurements provide an
intermediate compromise as they are deeper than the mean flux measurements,
while in principle having a more stable PSF than the stack images.

\citet{Tachibana18} show that the stack photometry works best when
morphologically classifying resolved, extended sources and unresolved point
sources. The methodology that we adopt here is extremely similar to
\citet{Tachibana18}, but we instead use PS1 forced photometry to classify
sources that do not have suitable stack photometry. The
forced-photometry-based model leads to slightly lower quality classifications
(see \S\ref{sec:results}).

\subsection{ML Training Set}\label{sec:training_set}

As a training set for the model, we use deep observations of the COSMOS field
from the \textit{Hubble Space Telescope} (HST). The superior
resolution of HST enables reliable morphological classifications for
sources as faint as $\sim$25\,mag \citep{Leauthaud07}. There are 80,867 bright
HST sources from \citet{Leauthaud07} that have PS1 counterparts
\citep[within a 1\arcsec match radius; see][]{Tachibana18} in the PS1
\texttt{ForcedMeanObject} table (see \S\ref{sec:features}). Of those, the
47,825 PS1 sources with $\mathtt{nDetections} \ge 1$ are adopted as the
training set for our model. This training set is $\sim$1.6\%
larger than the one used in \citet{Tachibana18} because more HST/COSMOS sources
are ``detected'' in PS1 forced photometry.\footnote{For this work a source is considered ``detected'' only if the \texttt{FPSFFlux}, \texttt{FPSFFluxErr}, \texttt{FKronFlux}, \texttt{FKronFluxErr}, \texttt{FApFlux}, \texttt{FApFluxErr} are all $> 0$ in at least one filter.}

\section{ML Model Features}\label{sec:ML_features}

\subsection{PS1 Forced Photometry Features}\label{sec:features}

Regardless of the choice of algorithm, the basic goal of a machine learning
model is to build a map between source features, numerical and/or categorical
properties that can be measured for an individual source, and labels, the
target output, often a classification, of the model. This mapping is learned
via a training set, a subset of the data with known labels, after which the
model can classify any source based on its features.

\citet{Tachibana18} introduced the concept of ``white flux'' features, whereby
measurements in the five individual PS1 filters were summed, via a weighted
mean, to produce a ``total'' flux or shape measurement across all
filters.\footnote{Only filters in which the source is detected are included in
the sum, see Equations~1 and 2 in \citet{Tachibana18}.} Machine learning
models are limited by their training sets: there is no guarantee that their
empirical mapping will correctly extend beyond the boundaries enclosed by the
training set. Given the significant systematic uncertainties associated with
Galactic reddening, and the tendency for spectroscopic samples, which are
typically used to define training sets, to be biased in their target selection
\citep[see e.g.,][]{Miller17}, the motivation for ``white flux'' features
becomes clear: they reduce potential biases in the final classifications due
to selection effects in how the training set sources were targeted. Therefore,
as in \citet{Tachibana18}, we use ``white flux'' features in this study.

The PS1 \texttt{StackObjectAttributes} table provides both flux and shape
(e.g., second moment of the radiation intensity) measurements in each of the
five PS1 filters, whereas the PS1 \texttt{ForcedMeanObject} table only
provides flux measurements.\footnote{The PS1 \texttt{ForcedMeanObject} table
provides average measurements across all epochs on which a PS1 source is
observed, and the average second moment of the radiation intensity is somewhat
meaningless as the orientation of the detector and observing conditions vary
image to image.}

To create the feature set for our machine learning model, we create ``white
flux'' features for the six different flux measurements available in the
\texttt{ForcedMeanObject} table\footnote{The original PS1 PSC and the PS1-STRM
catalogs are both constructed using the first PS1 data release. This study
uses measurements from the second PS1 data release, which corrects a
percent-level flat-field correction that was applied with the wrong sign in
DR1 \citep{Beck20}.} (\texttt{FPSFFlux}, \texttt{FKronFlux}, \texttt{FApFlux},
\texttt{FmeanflxR5}, \texttt{FmeanflxR6}, \texttt{FmeanflxR7}), as well as the
\texttt{E1} and \texttt{E2} measurements, which represent the mean
polarization parameters from \citet{Kaiser95}. We use flux ratios, rather than
the raw flux measurements, which provide morphological classifications that
are independent of S/N \citep{Lupton01}.

\begin{figure*}
    \centering
    \includegraphics[width=6.5in]{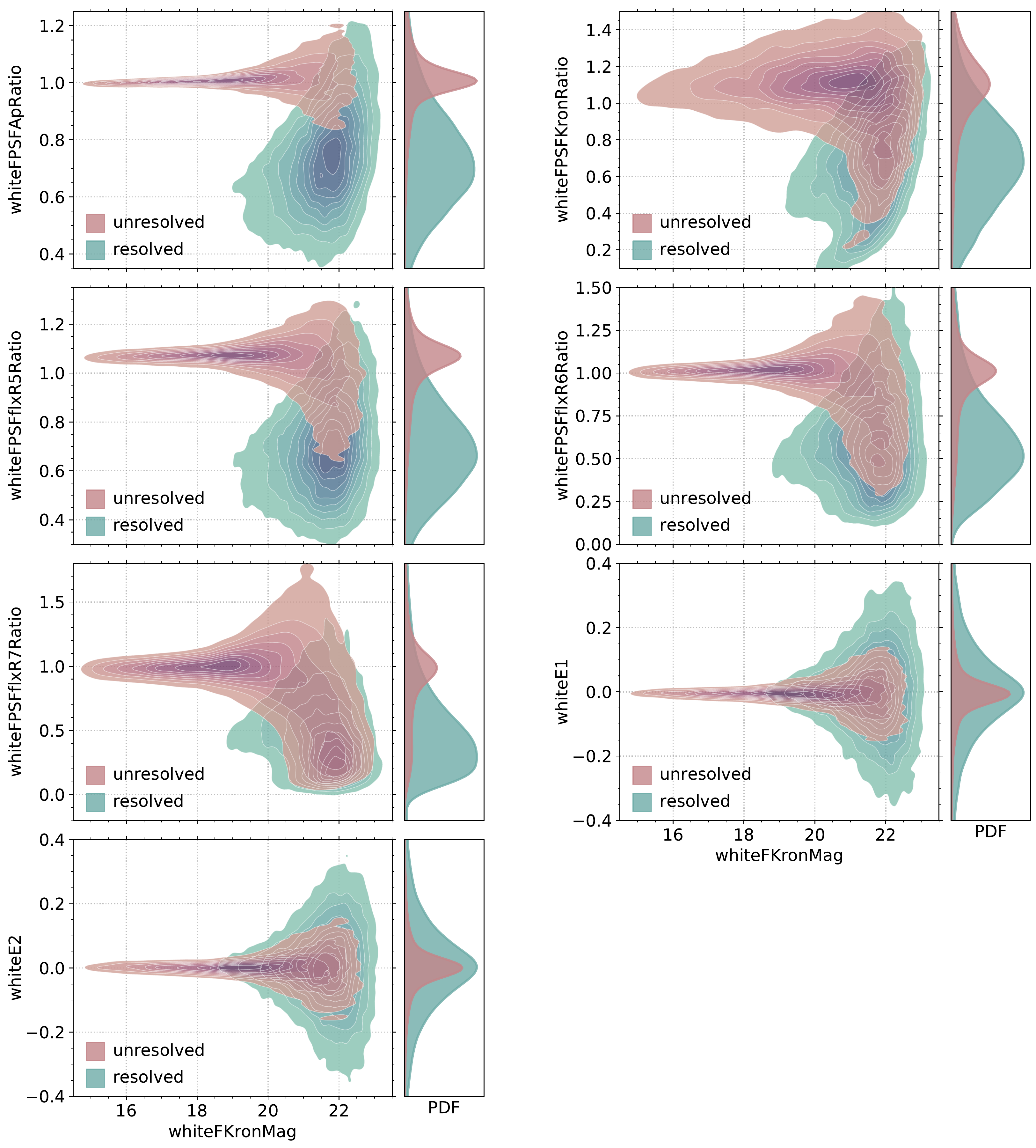}
    \caption{The primary square panels show Gaussian KDEs of the PDF for each
    of the ``white flux'' features as a function of \texttt{whiteFKronMag}
    ($=-2.5\log_{10}[\mathtt{whiteFKronFlux}/3631]$) for all sources in the
    training set. Unresolved point sources are shown via the red-purple
    contours, while resolved, extended objects are shown via blue-green
    contours. The shown contour levels extend from 0.9 to 0.1 in 0.1
    intervals. To the right of each primary panel is a marginalized 1D KDE of
    the PDF for the individual features, where the amplitudes of the KDEs have
    been normalized by the relative number of point sources and extended
    objects.}
    \label{fig:features}
\end{figure*} 

Our final model includes nine features, five flux ratios:
\begin{align*}
    \mathtt{whiteFPSFApRatio} &= \frac{\mathtt{whiteFPSFlux}}{\mathtt{whiteFApFlux}},\\
    \mathtt{whiteFPSFKronRatio} &= \frac{\mathtt{whiteFPSFlux}}{\mathtt{whiteFKronFlux}},\\
    \mathtt{whiteFPSFFmeanflxR5Ratio} &= \frac{\mathtt{whiteFPSFlux}}{\mathtt{whiteFmeanflxR5Flux}},\\
    \mathtt{whiteFPSFFmeanflxR6Ratio} &= \frac{\mathtt{whiteFPSFlux}}{\mathtt{whiteFmeanflxR6Flux}},\\
    \mathtt{whiteFPSFFmeanflxR7Ratio} &= \frac{\mathtt{whiteFPSFlux}}{\mathtt{whiteFmeanflxR7Flux}},
\end{align*}
the white polarization parameters: \texttt{whiteE1} and \texttt{whiteE2}, and
two ``simple'' distance measures: \texttt{whiteFPSFKronDist} and
\texttt{whiteFPSFKronDist} (see \S\ref{sec:simple_model}). The distribution of
these features for stars and galaxies in the training set is shown in
Figures~\ref{fig:features}, \ref{fig:psf_ap}, and \ref{fig:psf_kron}.

Figure~\ref{fig:features} shows that \texttt{whiteFPSFApRatio} is the most
useful feature, aside from the ``simple'' features, to separate resolved and
unresolved sources. This intuitively makes sense as PS1 \texttt{ApFlux}
measurements are matched to the seeing, whereas the \texttt{R5flx},
\texttt{R6flx}, \texttt{R7flx} measurements use fixed aperture sizes. With
multiple images taken under different observing conditions contributing to the
final forced flux measurements, fixed aperture measurements should be more
noisy.

\subsection{The ``Simple'' Distance Features}\label{sec:simple_model}

\citet{Tachibana18} introduced a ``simple'' model to classify sources based
solely on their measured \texttt{whitePSFFlux} and \texttt{whiteKronFlux}. The
model was inspired by the use of flux ratios, which have been shown to provide
a good discriminant between resolved and unresolved sources \citep[e.g., the
SDSS morphological \texttt{CLASS} parameter;][]{Lupton01}. At moderate to low
S/N, however, flux ratios no longer provide accurate classifications (see
e.g., Figure~\ref{fig:features}). The simple model from \citet{Tachibana18}
leverages this fact by measuring the distance of each source from a line drawn
in the \texttt{whitePSFFlux}--\texttt{whiteKronFlux} plane. Unlike a flux
ratio, the simple model preserves information about the S/N, meaning sources
with large absolute distances from the dividing line can be classified with
greater confidence.

Following from Equation~3 in \citet{Tachibana18}, ``simple'' features can be
calculated as:
\begin{equation}
 \mathtt{whiteF1F2Dist}(a) = 
 \frac{\mathtt{whiteF1} - a\times\mathtt{whiteF2}}{ \sqrt{1 + a^2}},
\end{equation}
where \texttt{whiteF1} and \texttt{whiteF2} are the ``white flux''
measurements introduced in \S\ref{sec:features} (e.g.,
\texttt{whiteFKronFlux}), $a$ is the slope of the line in the
\texttt{whiteF1}--\texttt{whiteF2} plane, and \texttt{whiteF1F2Dist} is the
orthogonal distance of a source from the line (sources above the line have
positive values). For this study we construct two simple features for
inclusion in our machine learning model: \texttt{whiteFPSFFKronDist} and
\texttt{whiteFPSFFApDist}.

\begin{figure}
    \centering
    \includegraphics[width=\columnwidth]{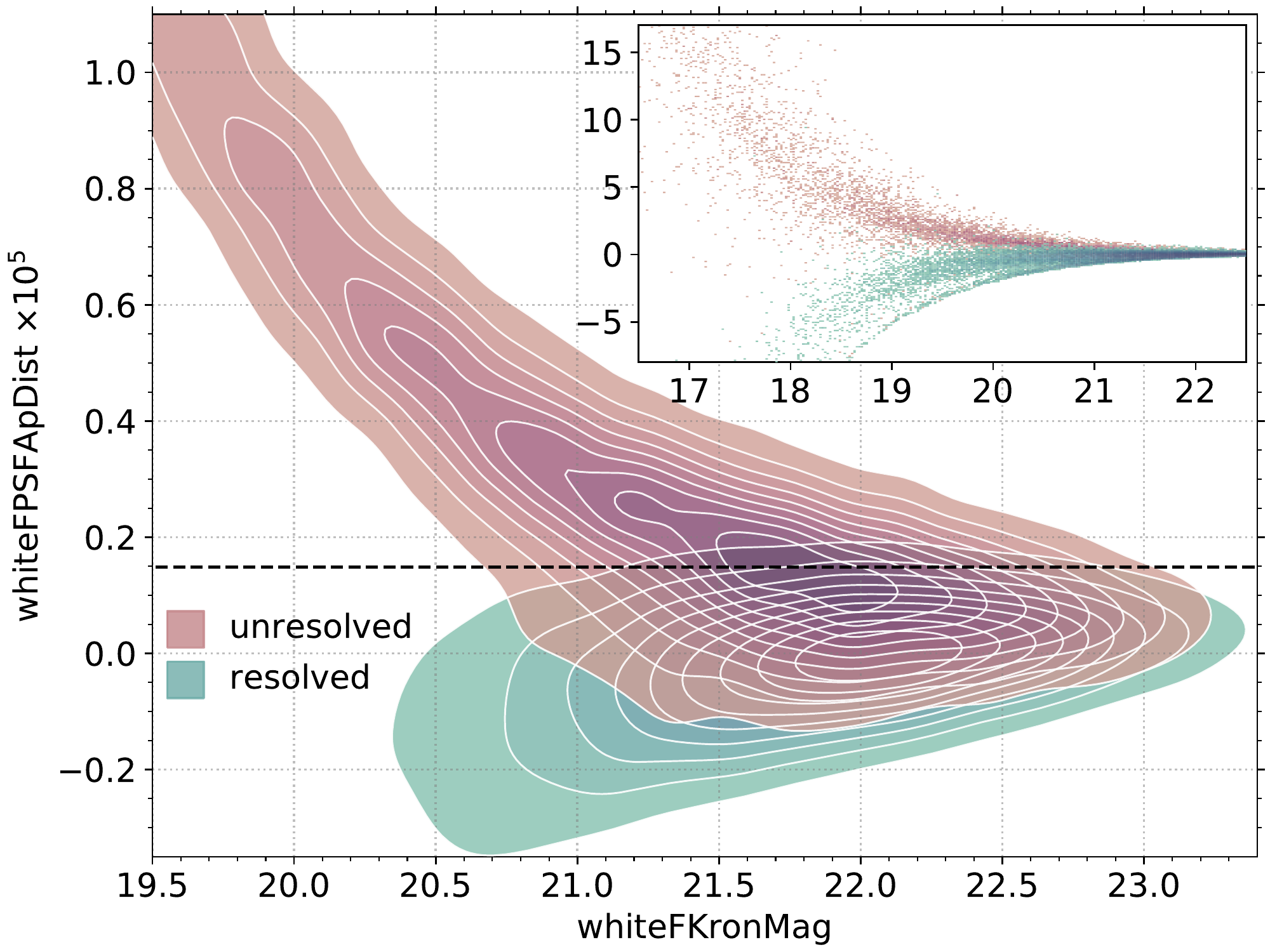}
    \caption{The distribution of $\mathtt{whiteFPSFKronDist}$ values for
    resolved, extended sources and unresolved point sources from the training
    set as a function of \texttt{whiteKronMag}. The colors and contours are
    the same as Figure~\ref{fig:features}. The horizontal dashed line shows
    the optimal threshold ($\mathtt{whiteFPSFKronDist} \ge 1.48 \times
    10^{-6}$) for resolved--unresolved classification. The upper-right inset
    shows a zoom-out highlighting the stark difference between stars and
    galaxies at the bright end.}
    \label{fig:psf_ap}
\end{figure}

\begin{figure}
    \centering
    \includegraphics[width=\columnwidth]{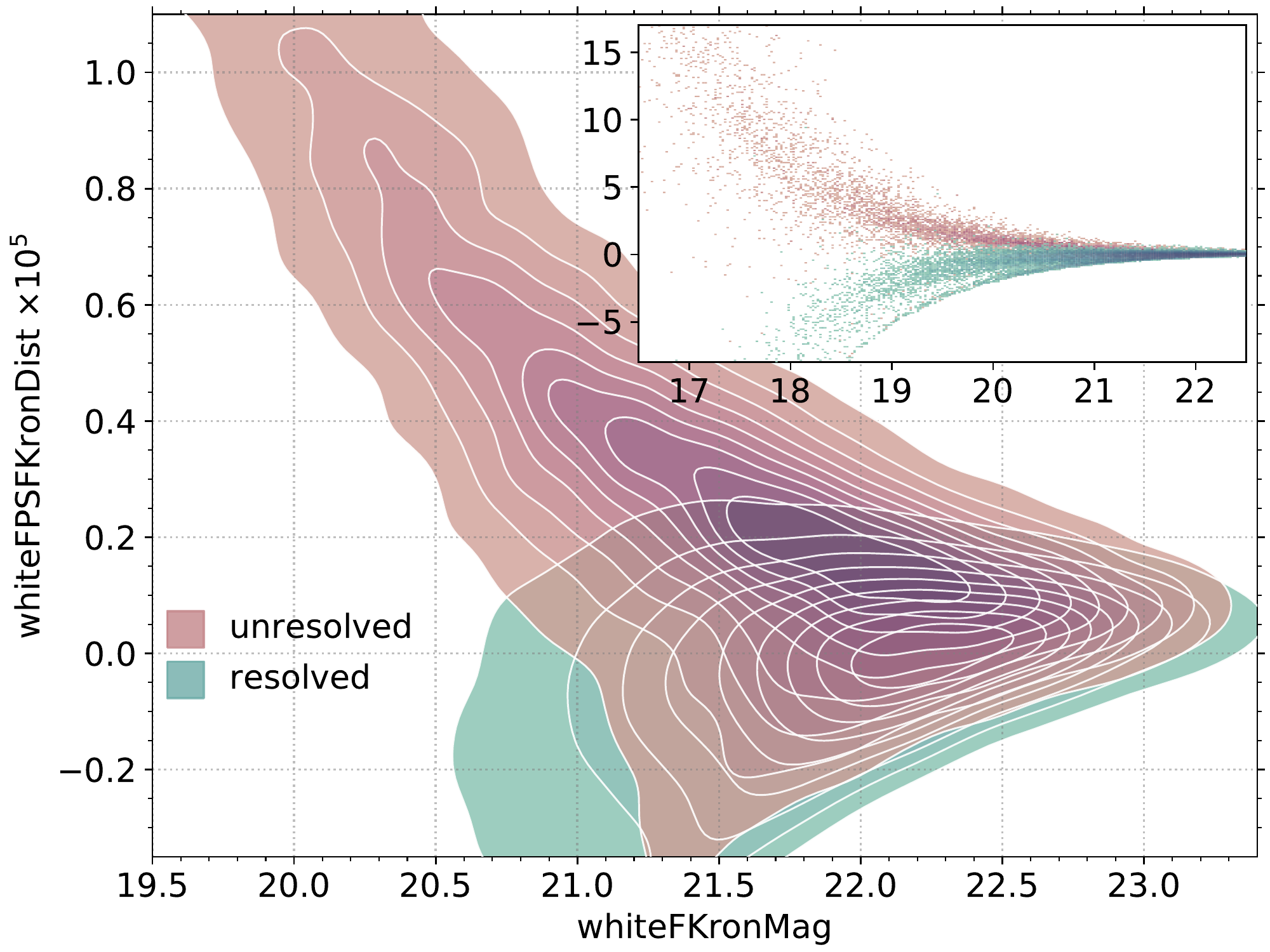}
    \caption{Same as Figure~\ref{fig:psf_ap}, but showing the distribution for
    \texttt{whiteFPSFKronDist}. A horizontal line is not shown as we do not
    recommend the use of only \texttt{whiteFPSFKronDist} for
    resolved-unresolved classification.}
    \label{fig:psf_kron}
\end{figure}

We determine the optimal value of $a$ for the simple features via cross
validation. We find $a = 0.7512$ for the \texttt{whiteFPSFFKronDist} feature
and $a = 0.7784$ for the \texttt{whiteFPSFFApDist} feature maximizes the FoM
(see \S\ref{sec:ML_model}). Empirically \texttt{whiteFPSFFApDist} is better at
separating resolved and unresolved sources than \texttt{whiteFPSFFKronDist},
and therefore the ``simple'' model, discussed below, is based on
\texttt{whiteFPSFFApDist}. The \texttt{whiteFPSFFApDist} and
\texttt{whiteFPSFFKronDist} distribution of resolved and unresolved sources is
shown in Figures~\ref{fig:psf_ap} and \ref{fig:psf_kron}, respectively.

\section{Training the ML Model}\label{sec:ML_model}

We construct a model to maximize the figure of merit (FoM) for our
morphological classification model. Our aim is to retain nearly all the
resolved, extended sources while excluding as many unresolved point sources as
possible. Thus, our FoM is defined as the true positive rate
(TPR)\footnote{$\mathrm{TPR} = \mathrm{TP}/(\mathrm{TP} + \mathrm{FP})$, where
TP is the total number of true positive classifications and FP is the number
of false positives.} at a fixed false positive rate
(FPR)\footnote{$\mathrm{FPR} = \mathrm{FP}/(\mathrm{FP}+\mathrm{TN})$, where
TN is the number of true negatives.} = 0.005.

Using the nine features from \S\ref{sec:ML_features}, we use the random forest
(RF) algorithm \citep{Breiman01}, as implemented in \texttt{scikit-learn}
\citep{Pedregosa11}, to classify PS1 sources as resolved or unresolved.
Briefly, the RF algorithm constructs an ensemble of decision trees
\citep{Breiman84}, where each tree is constructed using a bootstrapped sample
of the training set \citep[a method known as ``bagging'';][]{Breiman96} and
the split for each branch within the tree is selected from a random subset of
the full feature set. The result is a lower variance estimator than is
possible from a single decision tree.

To train the RF model, we replicate the procedure in \citet{Tachibana18}. We
use $k$-fold cross validation (CV) to optimize the model tuning parameters,
namely the number of trees in the forest $N_\mathrm{tree}$, the random number
of features for splitting at each node $m_\mathrm{try}$, and the minimum
number of sources in a terminal leaf of the tree $\mathtt{nodesize}$. Our CV
procedure utilizes both an inner and outer loop, each with $k = 10$ folds. In
the inner loop, a $k = 10$ folds CV grid search is performed over the three
tuning parameters, while predictions from the optimal grid location are
applied to the 1/10 of the training set that was withheld in the outer loop.
This process is then repeated for the remaining 9 folds in the outer loop. We
adopt the average results from the 10 different grid searches to arrive at
optimal model parameters of: $N_\mathrm{tree} = 900$, $m_\mathrm{try}
= 3$, and $\mathtt{nodesize} = 2$. The RF model results are not strongly
dependent on the final choice of tuning parameters.

\section{Results}\label{sec:results}

\subsection{Model Performance}

\begin{figure}[t]
 \centering
  \includegraphics[width=\columnwidth]{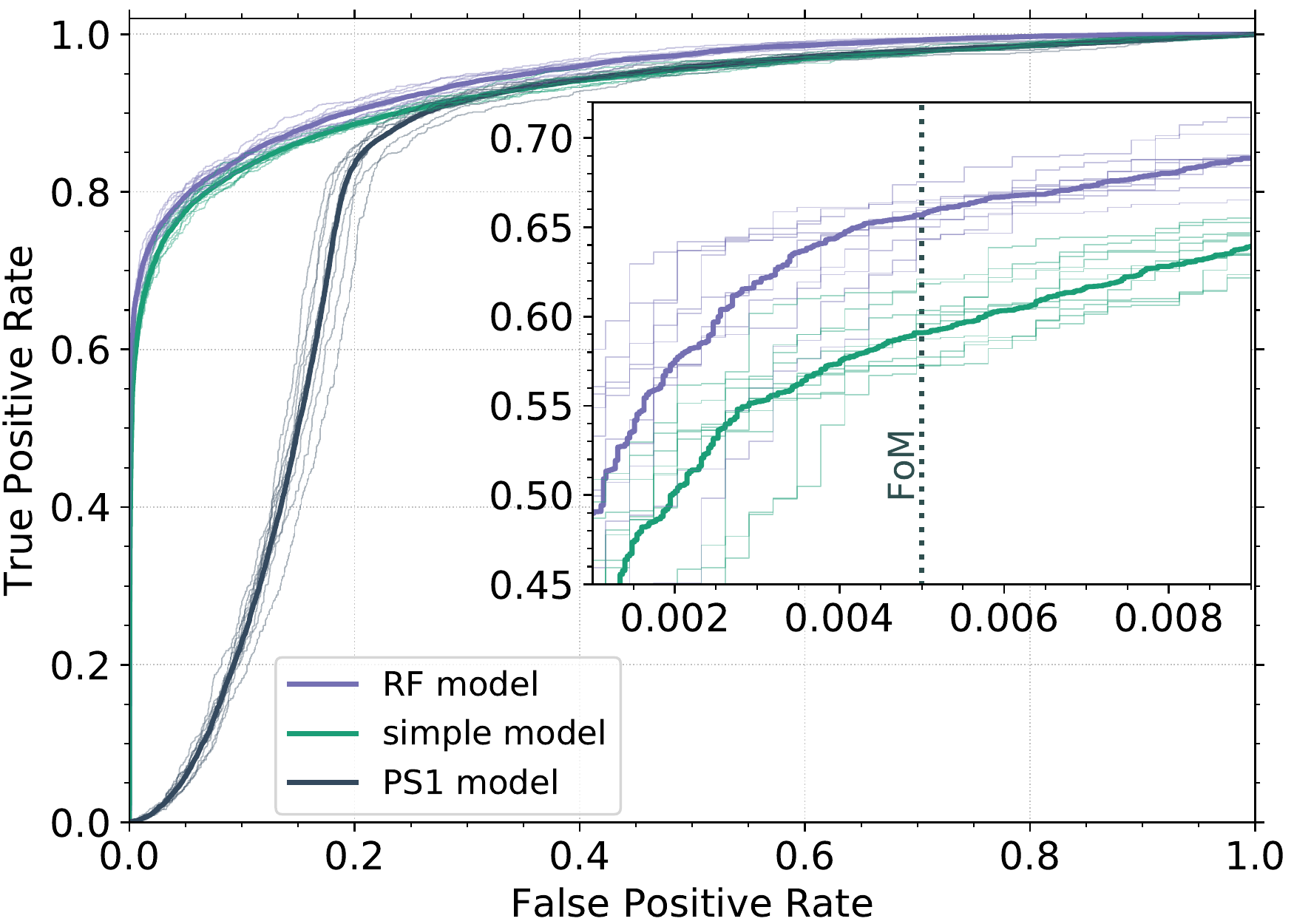}
  \caption{ ROC curves comparing the relative performance of the PS1, simple,
  and RF models for HST sources with $i_\mathrm{PS1}$ detections. The thick
  slate gray, green, and purple lines show the ROC curves for the PS1, simple,
  and RF models, respectively. The light, thin lines show the ROC curves for
  the individual CV folds. The inset on the right shows a zoom in around FPR =
  0.005, shown as a dotted vertical line, corresponding to the FoM (the PS1
  model is not shown in the inset, because it has very low FoM).}
  \label{fig:hst_roc}
\end{figure}

Our aim is to maximize the FoM of the RF model. We show receiver operating
characteristic (ROC) curves of the RF, simple, and PS1\footnote{The PS1 model
is defined by a single hard cut on the PSF--Kron flux ratio measured in the
$i_\mathrm{PS1}$ band \citep[for further details see][]{Tachibana18}.} models
in Figure~\ref{fig:hst_roc}. From Figure~\ref{fig:hst_roc}, it is clear that
the RF and simple models greatly outperform the PS1 model. Furthermore, while
the gains are modest, the inclusion of all the ``white flux'' features and use
of machine learning is justified as the RF model produces a higher FoM than
the simple model.

The FoM of each of the three models is summarized in Table~\ref{tbl:hst_cv}.
In addition to providing the largest FoM, the RF model is also the most
accurate and it has the largest area under the ROC curve (ROC AUC). We
robustly conclude that, of the models considered here, the RF model is best.
Comparing with Table~1 in \citet{Tachibana18}, we find that the
forced-photometry features derived in this study do not provide the same
discriminating power as the PS1 stack-photometry features used in
\citet{Tachibana18}. Our new model performs $\sim$7\% worse than the one in
\citet{Tachibana18}. In \S\ref{sec:ztf_pipeline}, we argue that this slight
reduction in performance is more than offset by the $\sim$144 million
additional sources that are now classified using the forced-photometry
features.

\begin{deluxetable}{cccc}
    \tablecolumns{4} 
    \tablewidth{0pt} 
    \tablecaption{ CV Results for the Training Set \label{tbl:hst_cv}}
    \tablehead{ 
    \colhead{model} & \colhead{FoM} & \colhead{Accuracy} & \colhead{ROC AUC}
    }
    \startdata
    RF & \textbf{0.657} $\pm$ 0.016 & \textbf{0.918} $\pm$ 0.003 & \textbf{0.945} $\pm$ 0.003 \\
    simple & 0.591 $\pm$ 0.017 & 0.910 $\pm$ 0.007 & 0.930 $\pm$ 0.003 \\
    PS1 & 0.002 $\pm$ 0.001 & 0.764 $\pm$ 0.011 & 0.827 $\pm$ 0.009 \\
    \enddata
    \tablecomments{Uncertainties represent the sample standard deviation for the 10 individual folds used in CV.}
\end{deluxetable}

\begin{figure}[t]
 \centering
  \includegraphics[width=\columnwidth]{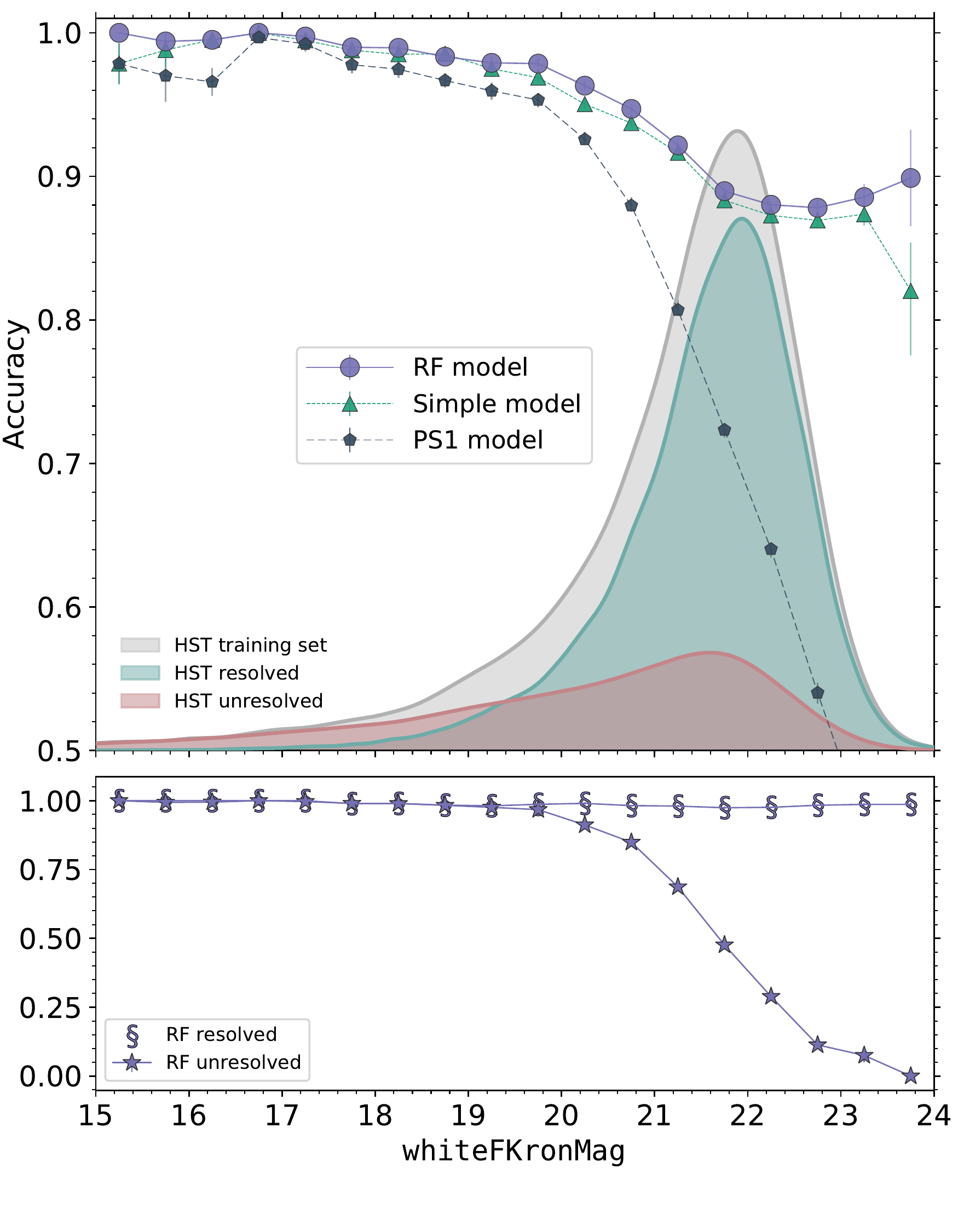}
  \caption{\textit{Top}: Model accuracy as a function of
  \texttt{whiteFKronMag} for HST sources with $i_\mathrm{PS1}$ detections.
  Accuracy curves for the PS1, simple and RF models are shown as slate gray
  pentagons, green triangles, and purple circles, respectively. The bin widths
  are 0.5\,mag, and the error bars represent the 68\% interval from bootstrap
  resampling. Additionally, a Gaussian KDE of the PDF for the training set, as
  well as the unresolved point sources and resolved, extended objects in the
  same subset is shown in the shaded gray, red, and green regions,
  respectively. The amplitude of the star and galaxy PDFs have been normalized
  by their relative ratio compared to the full $i_\mathrm{PS1}$-band subset.
  \textit{Bottom}: accuracy of resolved and unresolved classifications as a
  function of \texttt{whiteFKronMag} from the RF model (i.e., the TPR when
  treating each class as the positive class). Nearly all the resolved sources
  are correctly classified, because they dominate by number at low S/N (see
  text), while only bright unresolved sources are correctly classified.}
  \label{fig:hst_acc}
\end{figure}

We show the CV accuracy of the RF, simple, and PS1 models as a function of
\texttt{whiteFKronMag} in Figure~\ref{fig:hst_acc}. As in \citet{Tachibana18},
we find that the RF model provides more accurate classifications than the
alternatives.

The accuracy of each model shown in Figure~\ref{fig:hst_acc} decreases for
lower S/N sources. The accuracy curve for the RF and simple models feature a
slight departure from expectation in that they do not decrease much from 22 to
24\,mag. This quasi-plateau in the model accuracy can be understood as the
result of two components of the training set: (i) unresolved sources
completely dominate the source counts at these magnitudes, and (ii) the
well-defined locus of unresolved sources in the training set (see
Figure~\ref{fig:features}) becomes heavily blended with the resolved source
population at these brightness levels. Taken together the model will be biased
towards classifying all faint sources as resolved, despite the fact that we do
not explicitly include flux measurements in the feature set. With 88.5\% of
the $\mathtt{whiteFKronMag} > 22.5$\,mag training set sources being
unresolved, a quasi-plateau in accuracy of $\sim$88\% makes sense. This is
confirmed in the bottom panel of Figure~\ref{fig:hst_acc}, which shows the RF
model true positive rate (TPR) for both resolved and unresolved sources as a
function of \texttt{whiteFKronMag}. A near 100\% TPR for faint resolved
sources combined with a few correctly classified unresolved sources leads to
the observed quasi-plateau in Figure~\ref{fig:hst_acc}.

\subsection{The Updated PS1 PSC Catalog}\label{sec:ps1psc_update}

\begin{figure}
    \centering
    \includegraphics[width=\columnwidth]{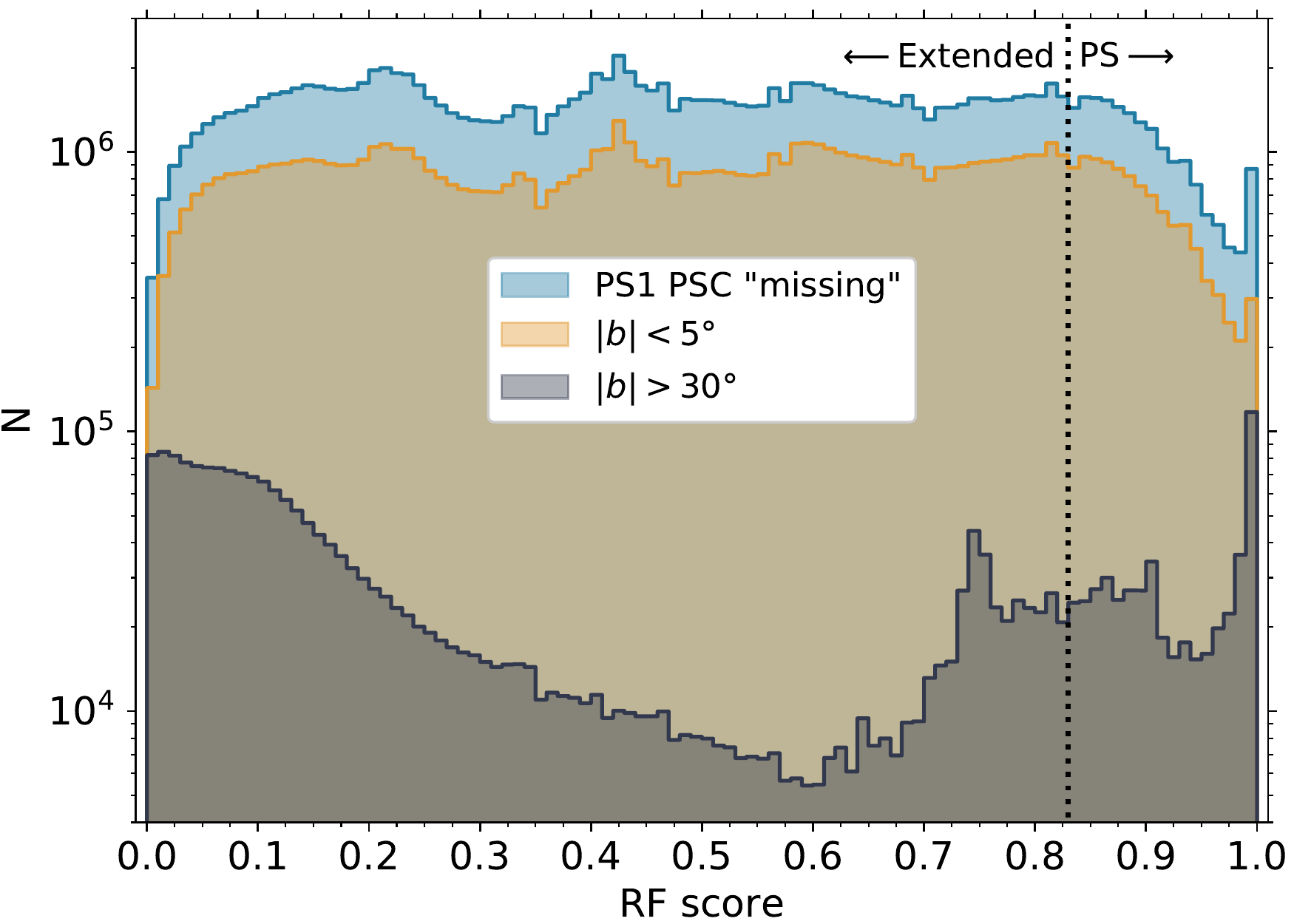}
    \caption{Histogram showing the RF classification scores for the $\sim$144
    million newly classified sources from PS1. All of the newly classified
    sources are shown in blue, while Galactic plane sources ($|b| <
    5^{\circ}$) are shown in orange, and high galactic latitude sources ($|b|
    > 30^{\circ}$) are shown in grey. The vertical dotted line shows the
    conservative classification threshold adopted in \citet{Tachibana18}
    (sources to the right of the line are considered point sources). The vast
    majority of the newly classified sources are in the Galactic plane.}
    \label{fig:psc_update}
\end{figure}

With a new RF model in hand, we can now provide morphological classifications
for the PS1 sources that are currently missing from the PS1 PSC. Of the
$\sim$426 million ``missing'' sources, $\sim$144 million have PS1 DR2
\texttt{ForcedMeanObject} photometry that pass our detection criteria (see
\ref{app:cat_counts} for more details). A histogram showing the distribution
of the RF classification score for these newly classified sources is shown in
Figure~\ref{fig:psc_update}.

Figure~\ref{fig:psc_update} shows that there are relatively few
high-confidence classifications (i.e., very likely extended sources with RF
score $\approx 0$ or very likely point sources with RF score $\approx 1$)
among the ``missing'' sources. Figure~\ref{fig:psc_update} also reveals the
likely explanation for this outcome: the vast majority of the newly classified
sources are in the Galactic plane. Of the $\sim$144 million newly classified
sources, $\sim$57\% have galactic latitude $\lvert b \rvert < 5$\,deg, while
$> 95$\% are in the Galactic plane ($\lvert b \rvert < 15$\,deg). The HST
COSMOS field, from which we derive our training set, has $b \approx 42$\,deg
and as a result includes very few stellar blends, which are common at low
galactic latitudes. The PS1 PSC also has significantly lower confidence
classifications in the Galactic plane \citep[see Figure~8 in][]{Tachibana18}.
That these sources were not ``detected'' in the PS1 stack images also suggests
that it is difficult to make reliable photometric measurements using the PS1
data, which could also contribute to the lower confidence classifications. The
upcoming third data release from the space-based \textit{Gaia} telescope
\citep{Perryman01} will improve this situation by classifying many of these
ambiguous sources as stars via parallax and proper motion measurements.

Ultimately, this update to the PS1 PSC has identified 17,945,494 likely point
sources using the optimized threshold from \citet[][RF score $\ge
0.83$]{Tachibana18}. While this number is small compared to the $\sim$734
million point sources in the original PS1 PSC, these $\sim$18 million newly
identified point sources would otherwise pass filters looking for
extragalactic transients in the ZTF alert stream. Their removal will reduce
the number of false positive transient candidates.

\section{Deployment in the ZTF Real-Time Pipeline}\label{sec:ztf_pipeline}

The ZTF real-time pipeline \citep{Masci19} provides AVRO alert packets
\citep[see][]{Patterson19} containing information (e.g., flux, position,
nearest neighbors) about any newly discovered sources of variability. The
packets include morphological classifications, based on the PS1 PSC
\citep{Tachibana18}, for the three closest closed sources in the ZTF
\texttt{Stars} table that are within 30\arcsec\ of the newly observed variable
source (see ~\ref{app:cat_counts} for a summary of the PS1 sources included in
the ZTF \texttt{Stars} table). There are $\sim$426 million PS1 sources in the
ZTF \texttt{Stars} table are not classified in the original PS1 PSC (see
\S\ref{sec:ps1psc_update}).

\subsection{Updating RF Classifications with Gaia Stars}\label{sec:gaia}

The \textit{Gaia} Early Data Release 3 includes high-precision astrometric
measurements collected over a 34 month timespan for $\sim$1.8 billion sources
\citep{Gaia-Collaboration20}. Within ZTF, the PS1 PSC is primarily used to
identify likely stars (i.e., point sources) and remove them from filters
searching for extragalactic transients. To that end, we can supplement the RF
classifications described in \S\ref{sec:ps1psc_update} with \textit{Gaia}
stars, which are identified via high-significance parallax and proper motion
detections.

A common threshold for determining ``high-significance'' is $\mathrm{S/N} \ge
5$, which in the case of gaussian uncertainties corresponds to a
$\sim$3$\times 10^{-7}$ probability that the observed signal is the result of
noise. We can therefore select stars from \textit{Gaia} sources with high S/N
parallax or proper motion measurements.\footnote{The total proper motion is
estimated by adding the proper motion in Right Ascension and Declination in
quadrature, see \citet{Tachibana18} for the corresponding uncertainty on this
quantity.} We adopt conservative significance thresholds because the formal
uncertainties from \textit{Gaia} are slightly underestimated
\citep{Fabricius20} and because most of the ``missing'' sources in the ZTF
\texttt{Stars} table are in the Galactic Plane (e.g.,
Figure~\ref{fig:psc_update}). \citet{Fabricius20} estimate that \textit{Gaia}
parallax measurements underestimate the uncertainties by a much as $\sim$60\%
in crowded regions. Similarly, proper motions are found to be underestimated
by as much $\sim$80\% in crowded regions \citep{Fabricius20}. We therefore
only consider \textit{Gaia} sources with a parallax $\mathrm{S/N} \ge 8$ or a
total proper motion $\mathrm{S/N} \ge 9$ to be stars.

Using the ESA \textit{Gaia}
archive\footnote{\url{https://gea.esac.esa.int/archive/}} we find there are
18,662,985 sources with either a high-significance parallax or proper motion
detection in the ZTF \texttt{Stars} table that lack a classification in the
original PS1 PSC. For these sources (11,479,512 of which have RF scores from
\S\ref{sec:ps1psc_update}) we update their scores to 1 in the ZTF
\texttt{Stars} table. This effectively excludes each of these sources from
filters designed to find extragalactic transients in the ZTF alert stream.

\subsection{Practical Implementation of the Updated Catalog}

Moving forward, ZTF alert packets now include $\sim$152 million additional
classifications ($\sim$133.6 million RF classifications from
\S\ref{sec:ps1psc_update}, and $\sim$18.6 million from
\S\ref{sec:gaia}.\footnote{This change will occur upon the journal acceptance
of this paper.} The addition of these new classifications to the ZTF AVRO
packets should not affect existing alert-stream filters, as we describe below.

\begin{deluxetable*}{l|lccccc}
    \tablecolumns{7}
    \tablewidth{0pt}
    \tablecaption{TPR and FPR for TM18 Thresholds\label{tbl:thresh}}
    \tablehead{
    \colhead{Catalog} & \colhead{Threshold} & \colhead{0.829} & \colhead{0.724} & \colhead{0.597} & \colhead{0.397} & \colhead{0.224}
    }
    \startdata
    \multirow{2}{*}{TM18} & TPR & 0.734 & 0.792 & 0.843 & 0.0904 & 0.947 \\
     & FPR & 0.005 & 0.01 & 0.02 & 0.05 & 0.1 \\
    \hline
    \multirow{2}{*}{This work} & TPR & 0.684$^{+0.005}_{-0.005}$ & 0.730$^{+0.004}_{-0.005}$ & 0.772$^{+0.004}_{-0.004}$ & 0.833$^{+0.004}_{-0.003}$ & 0.902$^{+0.004}_{-0.004}$\\
     & FPR & 0.005$^{+0.001}_{-0.000}$ & 0.009$^{+0.001}_{-0.001}$ & 0.016$^{+0.001}_{-0.001}$ & 0.041$^{+0.002}_{-0.001}$ & 0.111$^{+0.002}_{-0.003}$\\
    \enddata
    \tablecomments{The table reports the TPR and FPR for different 
    classification thresholds given in Table~3 in \citet{Tachibana18}. 
    To estimate the TPR and FPR we perform 10-fold CV on the entire training 
    set, but only include sources with $\mathtt{nDetections} \ge 3$ in the 
    final TPR and FPR calculations. The first row (TM18) summarizes the 
    results from \citet{Tachibana18}, while the second row uses the RF model 
    from this study. The reported uncertainties represent the central 90\% 
    interval from 100 bootstrap resamples of the training set.}
\end{deluxetable*}

While a one-to-one mapping of point-source classification scores cannot be
made between \citet{Tachibana18} and this study, the similarity between the
two methodologies leads to classifications that are highly similar.
Table~\ref{tbl:thresh} summarizes the TPR and FPR for different classification
thresholds using the model from \citet{Tachibana18} and the RF model created
in this study. The PS1 stack photometry used in \citet{Tachibana18}
consistently produces a higher TPR, by $\sim$6--8\%, than the PS1 forced
photometry. The PS1 forced photometry used in this study does have a lower FPR
than \citet{Tachibana18} for all but the most liberal point-source
classification cuts. Thus, applying classification cuts developed for the
original PS1 PSC will ultimately lead to a higher TPR, as previously
unclassified point sources can now be removed from the stream, without
experiencing an overall increase in the FPR. As a result, we conclude that the
vast majority of users will not experience any significant change in the
results to their filters, aside from a slight reduction in false negatives
(stars that are classified as galaxies), following the update to the ZTF
\texttt{Stars} table.

\section{Discussion}\label{sec:discussion}

During the preparation of this manuscript, \citet{Beck20} published the
Pan-STARRS1 Source Types and Redshifts with Machine learning (PS1-STRM)
catalog, which includes the machine learning classification of PS1 sources as
either stars, galaxies, or quasars. Like this study, \citet{Beck20} use PS1
forced photometry to provide classifications. There are a couple of
differences between the catalogs: the PS1-STRM classifies all $\sim$2.9
billion sources in the PS1 \texttt{ForcedMeanObject} table, while the updated
PS1 PSC only classifies $\sim$half that many sources.\footnote{We note that
the majority of the additional classifications in the PS1-STRM are
$\mathtt{nDetections} \le 2$ sources with low S/N photometry. These
classifications therefore have a lot more uncertainty than the sources that
are in common between the PS1 PSC and the PS1-STRM.} Another difference
between the two catalogs is that the PS1-STRM uses a neural-network
classifier, whereas the PS1 PSC uses the RF algorithm. Finally, the PS1-STRM
uses full color information in their classifier whereas the PS1 PSC uses
``white flux'' features (see \S\ref{sec:features}).

The most important distinction between the two catalogs, in our estimation, is
their training sets. The PS1-STRM is trained using spectroscopic labels that
predominantly come from the Sloan Digital Sky Survey
\citep[SDSS;][]{Abolfathi18}, whereas the PS1 PSC is trained via morphological
classifications from HST. An SDSS-based training set has two distinct
advantages: it is nearly two orders of magnitude larger than the HST training
set and it includes redshift information (which can be used to estimate
photometric redshifts, as is done in the PS1-STRM).

When considering only morphological classification, or similarly star-galaxy
separation, an SDSS-based training set produces biased classifications
\citep{Miller17,Tachibana18}. The SDSS spectroscopic targeting algorithm was
biased towards specific source classes, such as luminous red galaxies, and as
a result SDSS spectra are not representative of the average source in PS1
\citep[see Figure~1 in][]{Tachibana18}. Furthermore, the SDSS training set is
distinctly biased towards point sources at the faint end ($r \gtrsim
21$\,mag), which leads to models that overestimate the prevalence of point
sources at these brightness levels \citep[see e.g., Figure~7
in][]{Tachibana18}. It is for these reasons that we adopt the HST training set
for the PS1 PSC, despite its relatively modest size.

Ultimately, we recommend the use of both catalogs. Despite the different
methodologies and training sets, we expect the classifications to largely be in
agreement for bright sources ($r \lesssim 20$\,mag). In cases where the
catalogs agree, the classifications can be treated as extremely confident.
Most of the disagreements will occur at the faint end, where both catalogs
will provide noisier estimates. For faint sources where the catalogs disagree,
users should consider applying an additional prior based on the observed
source counts in the Universe \citep[e.g.,][]{Henrion11}. At high galactic
latitudes, nearly all the very faint sources are galaxies, while within the
Galactic plane nearly everything will be a star.

\section{Conclusions}

We have presented an update to the PS1 PSC \citep{Tachibana18}, by classifying
$\sim$144 million sources that were previously ``missing.'' The new
classifications are made using a new RF model that utilizes photometric and
shape features from the PS1 DR2 \texttt{ForcedMeanObject} table.

The training set and methodology are nearly identical to those used in
\citet{Tachibana18}, with the major difference being that that study used
features from the PS1 DR1 \texttt{StackObjectAttributes} table. The similarity
in methodology is intentional, as it allows new classifications for the
previously ``missing'' sources to be incorporated into the PS1 PSC without a
need for significant revisions to existing filters that are applied to the ZTF
alert stream. We find that the new model performs $\sim$6--8\% worse than the
one presented in \citet[][see Table~\ref{tbl:thresh}]{Tachibana18}.
Nevertheless, the slight degradation in performance is more than offset by the
addition of $>$144 million newly classified sources. The update to the PS1 PSC
presented here will improve the extragalactic transient search efficiency for
ZTF.

Spectroscopic observations from SDSS have now fueled the training sets for
machine learning models to separate stars and galaxies for more than a decade
\citep[e.g.,][]{Ball06,Beck20}. These labels have proven extremely valuable as
they have been applied to several surveys beyond SDSS
\citep[e.g.,][]{Miller17,Beck20}. Our ability to use methods built on
empirical training sets is going to be severely limited by the Vera C.\ Rubin
Observatory, whose images will be predominantly populated by extremely faint
sources \citep[$r \approx 24$\,mag;][]{Ivezic19}. With few spectroscopic
classifications of any kind at these depths, the separation of stars and
galaxies in Rubin Observatory data is going to largely rely on data from the
Rubin Observatory itself. In this regime machine learning is unlikely to play
a leading role, and purely photometric methods will be required to separate
stars and galaxies \citep[e.g.,][]{Slater20} and triage the Rubin Observatory
alert stream to remove stellar variables prior to the search for extragalactic
transients.

\acknowledgments

This work would not have been possible without the public release of the PS1
data. We thank F.~Masci and R.~Laher for helping us identify sources that were
not classified in the ZTF \texttt{Stars} table.

A.A.M.~is funded by the Large Synoptic Survey Telescope Corporation (LSSTC),
the Brinson Foundation, and the Moore Foundation in support of the LSSTC Data
Science Fellowship Program; he also receives support as a CIERA Fellow by the
CIERA Postdoctoral Fellowship Program (Center for Interdisciplinary
Exploration and Research in Astrophysics, Northwestern University). X.H.~is
supported by LSSTC, through Enabling Science Grant \#2020-01.

% \vspace{5mm}
\facilities{PS1 \citep{Chambers16}}

%% Similar to \facility{}, there is the optional \software command to allow 
%% authors a place to specify which programs were used during the creation of 
%% the manuscript. Authors should list each code and include either a
%% citation or url to the code inside ()s when available.

\software{\texttt{astropy} \citep{Astropy-Collaboration13,
Astropy-Collaboration18}, 
          \texttt{scipy} \citep{2020SciPy-NMeth},
          \texttt{matplotlib} \citep{Hunter07}, 
          \texttt{pandas} \citep{McKinney10},
          \texttt{scikit-learn} \citep{Pedregosa11}}

\appendix

\section{The ZTF--PS1 Morphological Catalog}\label{app:cat_counts}

The ZTF database contains a table (\texttt{Stars}) with sources selected from
the PS1 DR1 that are used to provide morphological classifications in the ZTF
alert packets. The ZTF \texttt{Stars} table was seeded from the PS1
\texttt{MeanObject} table and includes all PS1 \texttt{MeanObject} sources
with $\mathtt{nDetections} \ge 3$.\footnote{Immediately after the release of
PS1 DR1 it was recommended that sources detected on at least three individual
PS1 images were unlikely to be spurious. Hence, the use of this selection cut
for the ZTF \texttt{Stars} table.} There are 1,919,106,844 sources in the ZTF
\texttt{Stars} table. Of these, 1,484,281,394 are classified in the PS1 PSC
and another 8,520,167 are classified as point sources based on \textit{Gaia}
parallax and/or proper motion measurements \citep{Tachibana18}. Therefore,
there are 426,305,283 sources in the ZTF \texttt{Stars} table that did not
meet the quality cuts necessary to be included in the PS1 PSC.\footnote{Only
sources with a single row designated as the \texttt{primaryDetection} in the
PS1 \texttt{StackObjectAttributes} table and a stack ``detection''
\citep[i.e., the PSF, Kron, and aperture flux are all $>0$ in at least one
filter, see][]{Tachibana18} are included in the PS1 PSC. }

For the $\sim$426 million ZTF \texttt{Stars} table sources not in the PS1 PSC,
5,885,633 had multiple rows in the PS1 \texttt{StackObjectAttributes} table
with $\mathtt{primaryDetection} = 1$, while the rest were not ``detected'' in
the PS1 stacks. As described in \S\ref{sec:ps1psc_update}, 144,870,754 of the
previously ``missing'' sources pass our \texttt{ForcedMeanObject}
``detection'' criteria (see \S\ref{sec:training_set}) and are now included in
the PS1 PSC.

The remaining $\sim$281 million sources do not have reliable PS1 stack or
forced photometry, and as a result remain in the ZTF \texttt{Stars} table with
an ambiguous score of 0.5. About 8\% of the still unclassified ZTF
\texttt{Stars} table sources are not present in PS1 DR2 (mostly because they
have declination $\delta < -30$\,deg).\footnote{See
\url{https://outerspace.stsci.edu/display/PANSTARRS/PS1+DR2+caveats\#PS1DR2caveats-Missingdata} for more information.} Furthermore, $\sim$34\% of these
$\sim$281 million sources have $\mathtt{nDetections} = 3$, and $\sim$55\% have
$\mathtt{nDetections} \le 5$. That these sources have so few detections in PS1
increases the probability that they may be spurious, and even if they are not
spurious, they are otherwise very low S/N detections, which do not produce
highly confident classifications.

\bibliographystyle{aas_arxiv.bst}
\bibliography{/Users/adamamiller/Documents/tex_stuff/papers}

\begin{thebibliography}{}
\expandafter\ifx\csname natexlab\endcsname\relax\def\natexlab#1{#1}\fi
\providecommand{\url}[1]{\href{#1}{#1}}
\providecommand{\dodoi}[1]{}
\providecommand{\doarXiv}[1]{\href{https://arxiv.org/abs/#1}{\nolinkurl{https://arxiv.org/abs/#1}}}

\bibitem[{{Abbott} {et~al.}(2017){Abbott}, {Abbott}, {Abbott}, {Acernese},
  {Ackley}, {Adams}, {Adams}, {Addesso}, {Adhikari}, {Adya}, \&
  et~al.}]{Abbott17a}
{Abbott}, B.~P., {Abbott}, R., {Abbott}, T.~D., {et~al.} 2017,
  \href{http://dx.doi.org/10.3847/2041-8213/aa91c9}{\color{magenta}\apjl},
  \href{http://adsabs.harvard.edu/abs/2017ApJ...848L..12A}{\color{blue}848},
  \href{http://adsabs.harvard.edu/abs/2017ApJ...848L..12A}{\color{blue}L12}

\bibitem[{{Abolfathi} {et~al.}(2018){Abolfathi}, {Aguado}, {Aguilar}, {Allende
  Prieto}, {Almeida}, {Ananna}, {Anders}, {Anderson}, {Andrews}, {Anguiano},
  {Arag{\'o}n-Salamanca}, {Argudo-Fern{\'a}ndez}, {Armengaud}, {Ata},
  {Aubourg}, {Avila-Reese}, {Badenes}, {Bailey}, {Balland}, {Barger},
  {Barrera-Ballesteros}, {Bartosz}, {Bastien}, {Bates}, {Baumgarten},
  {Bautista}, {Beaton}, {Beers}, {Belfiore}, {Bender}, {Bernardi}, {Bershady},
  {Beutler}, {Bird}, {Bizyaev}, {Blanc}, {Blanton}, {Blomqvist}, {Bolton},
  {Boquien}, {Borissova}, {Bovy}, {Bradna Diaz}, {Brandt}, {Brinkmann},
  {Brownstein}, {Bundy}, {Burgasser}, {Burtin}, {Busca}, {Ca{\~n}as},
  {Cano-D{\'\i}az}, {Cappellari}, {Carrera}, {Casey}, {Cervantes Sodi}, {Chen},
  {Cherinka}, {Chiappini}, {Choi}, {Chojnowski}, {Chuang}, {Chung}, {Clerc},
  {Cohen}, {Comerford}, {Comparat}, {Correa do Nascimento}, {da Costa},
  {Cousinou}, {Covey}, {Crane}, {Cruz-Gonzalez}, {Cunha}, {da Silva Ilha},
  {Damke}, {Darling}, {Davidson}, {Dawson}, {de Icaza Lizaola}, {de la
  Macorra}, {de la Torre}, {De Lee}, {de Sainte Agathe}, {Deconto Machado},
  {Dell'Agli}, {Delubac}, {Diamond-Stanic}, {Donor}, {Downes}, {Drory}, {du Mas
  des Bourboux}, {Duckworth}, {Dwelly}, {Dyer}, {Ebelke}, {Davis Eigenbrot},
  {Eisenstein}, {Elsworth}, {Emsellem}, {Eracleous}, {Erfanianfar},
  {Escoffier}, {Fan}, {Fern{\'a}ndez Alvar}, {Fernandez-Trincado}, {Fernand o
  Cirolini}, {Feuillet}, {Finoguenov}, {Fleming}, {Font-Ribera}, {Freischlad},
  {Frinchaboy}, {Fu}, {G{\'o}mez Maqueo Chew}, {Galbany}, {Garc{\'\i}a
  P{\'e}rez}, {Garcia-Dias}, {Garc{\'\i}a-Hern{\'a}ndez}, {Garma Oehmichen},
  {Gaulme}, {Gelfand }, {Gil-Mar{\'\i}n}, {Gillespie}, {Goddard}, {Gonz{\'a}lez
  Hern{\'a}ndez}, {Gonzalez-Perez}, {Grabowski}, {Green}, {Grier}, {Gueguen},
  {Guo}, {Guy}, {Hagen}, {Hall}, {Harding}, {Hasselquist}, {Hawley}, {Hayes},
  {Hearty}, {Hekker}, {Hernand ez}, {Hernandez Toledo}, {Hogg},
  {Holley-Bockelmann}, {Holtzman}, {Hou}, {Hsieh}, {Hunt}, {Hutchinson},
  {Hwang}, {Jimenez Angel}, {Johnson}, {Jones}, {J{\"o}nsson}, {Jullo}, {Khan},
  {Kinemuchi}, {Kirkby}, {Kirkpatrick}, {Kitaura}, {Knapp}, {Kneib},
  {Kollmeier}, {Lacerna}, {Lane}, {Lang}, {Law}, {Le Goff}, {Lee}, {Li}, {Li},
  {Lian}, {Liang}, {Lima}, {Lin}, {Long}, {Lucatello}, {Lundgren}, {Mackereth},
  {MacLeod}, {Mahadevan}, {Maia}, {Majewski}, {Manchado}, {Maraston},
  {Mariappan}, {Marques-Chaves}, {Masseron}, {Masters}, {McDermid}, {McGreer},
  {Melendez}, {Meneses-Goytia}, {Merloni}, {Merrifield}, {Meszaros}, {Meza},
  {Minchev}, {Minniti}, {Mueller}, {Muller-Sanchez}, {Muna}, {Mu{\~n}oz},
  {Myers}, {Nair}, {Nand ra}, {Ness}, {Newman}, {Nichol}, {Nidever},
  {Nitschelm}, {Noterdaeme}, {O'Connell}, {Oelkers}, {Oravetz}, {Oravetz},
  {Ort{\'\i}z}, {Osorio}, {Pace}, {Padilla}, {Palanque-Delabrouille},
  {Palicio}, {Pan}, {Pan}, {Parikh}, {P{\^a}ris}, {Park}, {Peirani},
  {Pellejero-Ibanez}, {Penny}, {Percival}, {Perez-Fournon}, {Petitjean},
  {Pieri}, {Pinsonneault}, {Pisani}, {Prada}, {Prakash}, {Queiroz}, {Raddick},
  {Raichoor}, {Barboza Rembold}, {Richstein}, {Riffel}, {Riffel}, {Rix},
  {Robin}, {Rodr{\'\i}guez Torres}, {Rom{\'a}n-Z{\'u}{\~n}iga}, {Ross},
  {Rossi}, {Ruan}, {Ruggeri}, {Ruiz}, {Salvato}, {S{\'a}nchez}, {S{\'a}nchez},
  {Sanchez Almeida}, {S{\'a}nchez-Gallego}, {Santana Rojas}, {Santiago},
  {Schiavon}, {Schimoia}, {Schlafly}, {Schlegel}, {Schneider}, {Schuster},
  {Schwope}, {Seo}, {Serenelli}, {Shen}, {Shen}, {Shetrone}, {Shull}, {Silva
  Aguirre}, {Simon}, {Skrutskie}, {Slosar}, {Smethurst}, {Smith}, {Sobeck},
  {Somers}, {Souter}, {Souto}, {Spindler}, {Stark}, {Stassun}, {Steinmetz},
  {Stello}, {Storchi-Bergmann}, {Streblyanska}, {Stringfellow}, {Su{\'a}rez},
  {Sun}, {Szigeti}, {Taghizadeh-Popp}, {Talbot}, {Tang}, {Tao}, {Tayar},
  {Tembe}, {Teske}, {Thakar}, {Thomas}, {Tissera}, {Tojeiro}, {Tremonti},
  {Troup}, {Urry}, {Valenzuela}, {van den Bosch}, {Vargas-Gonz{\'a}lez},
  {Vargas-Maga{\~n}a}, {Vazquez}, {Villanova}, {Vogt}, {Wake}, {Wang},
  {Weaver}, {Weijmans}, {Weinberg}, {Westfall}, {Whelan}, {Wilcots}, {Wild},
  {Williams}, {Wilson}, {Wood-Vasey}, {Wylezalek}, {Xiao}, {Yan}, {Yang},
  {Ybarra}, {Y{\`e}che}, {Zakamska}, {Zamora}, {Zarrouk}, {Zasowski}, {Zhang},
  {Zhao}, {Zhao}, {Zheng}, {Zheng}, {Zhou}, {Zhu}, {Zinn}, \&
  {Zou}}]{Abolfathi18}
{Abolfathi}, B., {Aguado}, D.~S., {Aguilar}, G., {et~al.} 2018,
  \href{http://dx.doi.org/10.3847/1538-4365/aa9e8a}{\color{magenta}\apjs},
  \href{https://ui.adsabs.harvard.edu/abs/2018ApJS..235...42A}{\color{blue}235},
  \href{https://ui.adsabs.harvard.edu/abs/2018ApJS..235...42A}{\color{blue}42}

\bibitem[{{Astropy Collaboration} {et~al.}(2013){Astropy Collaboration},
  {Robitaille}, {Tollerud}, {Greenfield}, {Droettboom}, {Bray}, {Aldcroft},
  {Davis}, {Ginsburg}, {Price-Whelan}, {Kerzendorf}, {Conley}, {Crighton},
  {Barbary}, {Muna}, {Ferguson}, {Grollier}, {Parikh}, {Nair}, {Unther},
  {Deil}, {Woillez}, {Conseil}, {Kramer}, {Turner}, {Singer}, {Fox}, {Weaver},
  {Zabalza}, {Edwards}, {Azalee Bostroem}, {Burke}, {Casey}, {Crawford},
  {Dencheva}, {Ely}, {Jenness}, {Labrie}, {Lim}, {Pierfederici}, {Pontzen},
  {Ptak}, {Refsdal}, {Servillat}, \& {Streicher}}]{Astropy-Collaboration13}
{Astropy Collaboration}, {Robitaille}, T.~P., {Tollerud}, E.~J., {et~al.} 2013,
  \href{http://dx.doi.org/10.1051/0004-6361/201322068}{\color{magenta}\aap},
  \href{http://adsabs.harvard.edu/abs/2013A%26A...558A..33A}{\color{blue}558},
  \href{http://adsabs.harvard.edu/abs/2013A%26A...558A..33A}{\color{blue}A33}

\bibitem[{{Astropy Collaboration} {et~al.}(2018){Astropy Collaboration},
  {Price-Whelan}, {Sip{\H{o}}cz}, {G{\"u}nther}, {Lim}, {Crawford}, {Conseil},
  {Shupe}, {Craig}, {Dencheva}, {Ginsburg}, {Vand erPlas}, {Bradley},
  {P{\'e}rez-Su{\'a}rez}, {de Val-Borro}, {Aldcroft}, {Cruz}, {Robitaille},
  {Tollerud}, {Ardelean}, {Babej}, {Bach}, {Bachetti}, {Bakanov}, {Bamford},
  {Barentsen}, {Barmby}, {Baumbach}, {Berry}, {Biscani}, {Boquien}, {Bostroem},
  {Bouma}, {Brammer}, {Bray}, {Breytenbach}, {Buddelmeijer}, {Burke},
  {Calderone}, {Cano Rodr{\'\i}guez}, {Cara}, {Cardoso}, {Cheedella}, {Copin},
  {Corrales}, {Crichton}, {D'Avella}, {Deil}, {Depagne}, {Dietrich}, {Donath},
  {Droettboom}, {Earl}, {Erben}, {Fabbro}, {Ferreira}, {Finethy}, {Fox},
  {Garrison}, {Gibbons}, {Goldstein}, {Gommers}, {Greco}, {Greenfield},
  {Groener}, {Grollier}, {Hagen}, {Hirst}, {Homeier}, {Horton}, {Hosseinzadeh},
  {Hu}, {Hunkeler}, {Ivezi{\'c}}, {Jain}, {Jenness}, {Kanarek}, {Kendrew},
  {Kern}, {Kerzendorf}, {Khvalko}, {King}, {Kirkby}, {Kulkarni}, {Kumar},
  {Lee}, {Lenz}, {Littlefair}, {Ma}, {Macleod}, {Mastropietro}, {McCully},
  {Montagnac}, {Morris}, {Mueller}, {Mumford}, {Muna}, {Murphy}, {Nelson},
  {Nguyen}, {Ninan}, {N{\"o}the}, {Ogaz}, {Oh}, {Parejko}, {Parley}, {Pascual},
  {Patil}, {Patil}, {Plunkett}, {Prochaska}, {Rastogi}, {Reddy Janga},
  {Sabater}, {Sakurikar}, {Seifert}, {Sherbert}, {Sherwood-Taylor}, {Shih},
  {Sick}, {Silbiger}, {Singanamalla}, {Singer}, {Sladen}, {Sooley},
  {Sornarajah}, {Streicher}, {Teuben}, {Thomas}, {Tremblay}, {Turner},
  {Terr{\'o}n}, {van Kerkwijk}, {de la Vega}, {Watkins}, {Weaver}, {Whitmore},
  {Woillez}, {Zabalza}, \& {Astropy Contributors}}]{Astropy-Collaboration18}
{Astropy Collaboration}, {Price-Whelan}, A.~M., {Sip{\H{o}}cz}, B.~M., {et~al.}
  2018, \href{http://dx.doi.org/10.3847/1538-3881/aabc4f}{\color{magenta}\aj},
  \href{https://ui.adsabs.harvard.edu/abs/2018AJ....156..123A}{\color{blue}156},
  \href{https://ui.adsabs.harvard.edu/abs/2018AJ....156..123A}{\color{blue}123}

\bibitem[{{Ball} {et~al.}(2006){Ball}, {Brunner}, {Myers}, \&
  {Tcheng}}]{Ball06}
{Ball}, N.~M., {Brunner}, R.~J., {Myers}, A.~D., \& {Tcheng}, D. 2006,
  \href{http://dx.doi.org/10.1086/507440}{\color{magenta}\apj},
  \href{http://adsabs.harvard.edu/abs/2006ApJ...650..497B}{\color{blue}650},
  \href{http://adsabs.harvard.edu/abs/2006ApJ...650..497B}{\color{blue}497}

\bibitem[{{Beck} {et~al.}(2020){Beck}, {Szapudi}, {Flewelling}, {Holmberg},
  {Magnier}, \& {Chambers}}]{Beck20}
{Beck}, R., {Szapudi}, I., {Flewelling}, H., {et~al.} 2020,
  \href{http://dx.doi.org/10.1093/mnras/staa2587}{\color{magenta}\mnras}

\bibitem[{{Bellm} {et~al.}(2019){Bellm}, {Kulkarni}, {Graham}, {Dekany},
  {Smith}, {Riddle}, {Masci}, {Helou}, {Prince}, {Adams}, {Barbarino},
  {Barlow}, {Bauer}, {Beck}, {Belicki}, {Biswas}, {Blagorodnova}, {Bodewits},
  {Bolin}, {Brinnel}, {Brooke}, {Bue}, {Bulla}, {Burruss}, {Cenko}, {Chang},
  {Connolly}, {Coughlin}, {Cromer}, {Cunningham}, {De}, {Delacroix}, {Desai},
  {Duev}, {Eadie}, {Farnham}, {Feeney}, {Feindt}, {Flynn}, {Franckowiak},
  {Frederick}, {Fremling}, {Gal-Yam}, {Gezari}, {Giomi}, {Goldstein},
  {Golkhou}, {Goobar}, {Groom}, {Hacopians}, {Hale}, {Henning}, {Ho}, {Hover},
  {Howell}, {Hung}, {Huppenkothen}, {Imel}, {Ip}, {Ivezi{\'c}}, {Jackson},
  {Jones}, {Juric}, {Kasliwal}, {Kaspi}, {Kaye}, {Kelley}, {Kowalski},
  {Kramer}, {Kupfer}, {Landry}, {Laher}, {Lee}, {Lin}, {Lin}, {Lunnan},
  {Giomi}, {Mahabal}, {Mao}, {Miller}, {Monkewitz}, {Murphy}, {Ngeow},
  {Nordin}, {Nugent}, {Ofek}, {Patterson}, {Penprase}, {Porter}, {Rauch},
  {Rebbapragada}, {Reiley}, {Rigault}, {Rodriguez}, {van Roestel}, {Rusholme},
  {van Santen}, {Schulze}, {Shupe}, {Singer}, {Soumagnac}, {Stein}, {Surace},
  {Sollerman}, {Szkody}, {Taddia}, {Terek}, {Van Sistine}, {van Velzen},
  {Vestrand}, {Walters}, {Ward}, {Ye}, {Yu}, {Yan}, \& {Zolkower}}]{Bellm19}
{Bellm}, E.~C., {Kulkarni}, S.~R., {Graham}, M.~J., {et~al.} 2019,
  \href{http://dx.doi.org/10.1088/1538-3873/aaecbe}{\color{magenta}\pasp},
  \href{https://ui.adsabs.harvard.edu/\#abs/2019PASP..131a8002B}{\color{blue}131},
  \href{https://ui.adsabs.harvard.edu/\#abs/2019PASP..131a8002B}{\color{blue}018002}

\bibitem[{{Bloom} {et~al.}(2012){Bloom}, {Richards}, {Nugent}, {Quimby},
  {Kasliwal}, {Starr}, {Poznanski}, {Ofek}, {Cenko}, {Butler}, {Kulkarni},
  {Gal-Yam}, \& {Law}}]{Bloom12}
{Bloom}, J.~S., {Richards}, J.~W., {Nugent}, P.~E., {et~al.} 2012,
  \href{http://dx.doi.org/10.1086/668468}{\color{magenta}\pasp},
  \href{http://adsabs.harvard.edu/abs/2012PASP..124.1175B}{\color{blue}124},
  \href{http://adsabs.harvard.edu/abs/2012PASP..124.1175B}{\color{blue}1175}

\bibitem[{Breiman(1996)}]{Breiman96}
Breiman, L. 1996,
  \href{http://dx.doi.org/10.1007/BF00058655}{\color{magenta}Machine Learning},
  24, 123

\bibitem[{Breiman(2001)}]{Breiman01}
---. 2001,
  \href{http://dx.doi.org/10.1023/A:1010933404324}{\color{magenta}Machine
  Learning}, 45, 5

\bibitem[{Breiman {et~al.}(1984)Breiman, Friedman, Stone, \&
  Olshen}]{Breiman84}
Breiman, L., Friedman, J., Stone, C.~J., \& Olshen, R.~A. 1984, Classification
  and regression trees (CRC press)

\bibitem[{{Brink} {et~al.}(2013){Brink}, {Richards}, {Poznanski}, {Bloom},
  {Rice}, {Negahban}, \& {Wainwright}}]{Brink13}
{Brink}, H., {Richards}, J.~W., {Poznanski}, D., {et~al.} 2013,
  \href{http://dx.doi.org/10.1093/mnras/stt1306}{\color{magenta}\mnras},
  \href{http://adsabs.harvard.edu/abs/2013MNRAS.435.1047B}{\color{blue}435},
  \href{http://adsabs.harvard.edu/abs/2013MNRAS.435.1047B}{\color{blue}1047}

\bibitem[{{Chambers} {et~al.}(2016){Chambers}, {Magnier}, {Metcalfe},
  {Flewelling}, {Huber}, {Waters}, {Denneau}, {Draper}, {Farrow}, {Finkbeiner},
  {Holmberg}, {Koppenhoefer}, {Price}, {Saglia}, {Schlafly}, {Smartt},
  {Sweeney}, {Wainscoat}, {Burgett}, {Grav}, {Heasley}, {Hodapp}, {Jedicke},
  {Kaiser}, {Kudritzki}, {Luppino}, {Lupton}, {Monet}, {Morgan}, {Onaka},
  {Stubbs}, {Tonry}, {Banados}, {Bell}, {Bender}, {Bernard}, {Botticella},
  {Casertano}, {Chastel}, {Chen}, {Chen}, {Cole}, {Deacon}, {Frenk},
  {Fitzsimmons}, {Gezari}, {Goessl}, {Goggia}, {Goldman}, {Grebel}, {Hambly},
  {Hasinger}, {Heavens}, {Heckman}, {Henderson}, {Henning}, {Holman}, {Hopp},
  {Ip}, {Isani}, {Keyes}, {Koekemoer}, {Kotak}, {Long}, {Lucey}, {Liu},
  {Martin}, {McLean}, {Morganson}, {Murphy}, {Nieto-Santisteban}, {Norberg},
  {Peacock}, {Pier}, {Postman}, {Primak}, {Rae}, {Rest}, {Riess}, {Riffeser},
  {Rix}, {Roser}, {Schilbach}, {Schultz}, {Scolnic}, {Szalay}, {Seitz},
  {Shiao}, {Small}, {Smith}, {Soderblom}, {Taylor}, {Thakar}, {Thiel},
  {Thilker}, {Urata}, {Valenti}, {Walter}, {Watters}, {Werner}, {White},
  {Wood-Vasey}, \& {Wyse}}]{Chambers16}
{Chambers}, K.~C., {Magnier}, E.~A., {Metcalfe}, N., {et~al.} 2016, ArXiv
  e-prints

\bibitem[{{De} {et~al.}(2020){De}, {Kasliwal}, {Tzanidakis}, {Fremling},
  {Adams}, {Andreoni}, {Bagdasaryan}, {Bellm}, {Bildsten}, {Cannella}, {Cook},
  {Delacroix}, {Drake}, {Duev}, {Dugas}, {Frederick}, {Gal-Yam}, {Goldstein},
  {Golkhou}, {Graham}, {Hale}, {Hankins}, {Helou}, {Ho}, {Irani}, {Jencson},
  {Kaye}, {Kulkarni}, {Kupfer}, {Laher}, {Leadbeater}, {Lunnan}, {Masci},
  {Miller}, {Neill}, {Ofek}, {Perley}, {Polin}, {Prince}, {Quataert}, {Reiley},
  {Riddle}, {Rusholme}, {Sharma}, {Shupe}, {Sollerman}, {Tartaglia}, {Walters},
  {Yan}, \& {Yao}}]{De20}
{De}, K., {Kasliwal}, M.~M., {Tzanidakis}, A., {et~al.} 2020,
  \href{https://arxiv.org/abs/2004.09029}{\color{magenta}arXiv},
  \href{https://ui.adsabs.harvard.edu/abs/2020arXiv200409029D}{\color{blue}arXiv:2004.09029}

\bibitem[{{Drout} {et~al.}(2014){Drout}, {Chornock}, {Soderberg}, {Sand ers},
  {McKinnon}, {Rest}, {Foley}, {Milisavljevic}, {Margutti}, {Berger},
  {Calkins}, {Fong}, {Gezari}, {Huber}, {Kankare}, {Kirshner}, {Leibler},
  {Lunnan}, {Mattila}, {Marion}, {Narayan}, {Riess}, {Roth}, {Scolnic},
  {Smartt}, {Tonry}, {Burgett}, {Chambers}, {Hodapp}, {Jedicke}, {Kaiser},
  {Magnier}, {Metcalfe}, {Morgan}, {Price}, \& {Waters}}]{Drout14}
{Drout}, M.~R., {Chornock}, R., {Soderberg}, A.~M., {et~al.} 2014,
  \href{http://dx.doi.org/10.1088/0004-637X/794/1/23}{\color{magenta}\apj},
  \href{https://ui.adsabs.harvard.edu/abs/2014ApJ...794...23D}{\color{blue}794},
  \href{https://ui.adsabs.harvard.edu/abs/2014ApJ...794...23D}{\color{blue}23}

\bibitem[{{Duev} {et~al.}(2019){Duev}, {Mahabal}, {Masci}, {Graham},
  {Rusholme}, {Walters}, {Karmarkar}, {Frederick}, {Kasliwal}, {Rebbapragada},
  \& {Ward}}]{Duev19}
{Duev}, D.~A., {Mahabal}, A., {Masci}, F.~J., {et~al.} 2019,
  \href{http://dx.doi.org/10.1093/mnras/stz2357}{\color{magenta}\mnras},
  \href{https://ui.adsabs.harvard.edu/abs/2019MNRAS.489.3582D}{\color{blue}489},
  \href{https://ui.adsabs.harvard.edu/abs/2019MNRAS.489.3582D}{\color{blue}3582}

\bibitem[{{Fabricius} {et~al.}(2020){Fabricius}, {Luri}, {Arenou}, {Babusiaux},
  {Helmi}, {Muraveva}, {Reyl{\'e}}, {Spoto}, {Vallenari}, {Antoja}, {Balbinot},
  {Barache}, {Bauchet}, {Bragaglia}, {Busonero}, {Cantat-Gaudin}, {Carrasco},
  {Diakit{\'e}}, {Fabrizio}, {Figueras}, {Garcia-Gutierrez}, {Garofalo},
  {Jordi}, {Kervella}, {Khanna}, {Leclerc}, {Licata}, {Lambert}, {Marrese},
  {Masip}, {Ramos}, {Robichon}, {Robin}, {Romero-G{\'o}mez}, {Rubele}, \&
  {Weiler}}]{Fabricius20}
{Fabricius}, C., {Luri}, X., {Arenou}, F., {et~al.} 2020,
  \href{https://arxiv.org/abs/2012.06242}{\color{magenta}arXiv},
  \href{https://ui.adsabs.harvard.edu/abs/2020arXiv201206242F}{\color{blue}arXiv:2012.06242}

\bibitem[{{Fremling} {et~al.}(2020){Fremling}, {Miller}, {Sharma}, {Dugas},
  {Perley}, {Taggart}, {Sollerman}, {Goobar}, {Graham}, {Neill}, {Nordin},
  {Rigault}, {Walters}, {Andreoni}, {Bagdasaryan}, {Belicki}, {Cannella},
  {Bellm}, {Cenko}, {De}, {Dekany}, {Frederick}, {Golkhou}, {Graham}, {Helou},
  {Ho}, {Kasliwal}, {Kupfer}, {Laher}, {Mahabal}, {Masci}, {Riddle},
  {Rusholme}, {Schulze}, {Shupe}, {Smith}, {Velzen}, {Yan}, {Yao}, {Zhuang}, \&
  {Kulkarni}}]{Fremling20}
{Fremling}, C., {Miller}, A.~A., {Sharma}, Y., {et~al.} 2020,
  \href{http://dx.doi.org/10.3847/1538-4357/ab8943}{\color{magenta}\apj},
  \href{https://ui.adsabs.harvard.edu/abs/2020ApJ...895...32F}{\color{blue}895},
  \href{https://ui.adsabs.harvard.edu/abs/2020ApJ...895...32F}{\color{blue}32}

\bibitem[{{Gaia Collaboration} {et~al.}(2020){Gaia Collaboration}, {Brown},
  {Vallenari}, {Prusti}, {de Bruijne}, {Babusiaux}, \&
  {Biermann}}]{Gaia-Collaboration20}
{Gaia Collaboration}, {Brown}, A.~G.~A., {Vallenari}, A., {et~al.} 2020,
  \href{https://arxiv.org/abs/2012.01533}{\color{magenta}arXiv},
  \href{https://ui.adsabs.harvard.edu/abs/2020arXiv201201533G}{\color{blue}arXiv:2012.01533}

\bibitem[{{Gal-Yam} {et~al.}(2014){Gal-Yam}, {Arcavi}, {Ofek}, {Ben-Ami},
  {Cenko}, {Kasliwal}, {Cao}, {Yaron}, {Tal}, {Silverman}, {Horesh}, {De Cia},
  {Taddia}, {Sollerman}, {Perley}, {Vreeswijk}, {Kulkarni}, {Nugent},
  {Filippenko}, \& {Wheeler}}]{Gal-Yam14}
{Gal-Yam}, A., {Arcavi}, I., {Ofek}, E.~O., {et~al.} 2014,
  \href{http://dx.doi.org/10.1038/nature13304}{\color{magenta}\nat},
  \href{http://adsabs.harvard.edu/abs/2014Natur.509..471G}{\color{blue}509},
  \href{http://adsabs.harvard.edu/abs/2014Natur.509..471G}{\color{blue}471}

\bibitem[{{Gezari} {et~al.}(2012){Gezari}, {Chornock}, {Rest}, {Huber},
  {Forster}, {Berger}, {Challis}, {Neill}, {Martin}, {Heckman}, {Lawrence},
  {Norman}, {Narayan}, {Foley}, {Marion}, {Scolnic}, {Chomiuk}, {Soderberg},
  {Smith}, {Kirshner}, {Riess}, {Smartt}, {Stubbs}, {Tonry}, {Wood-Vasey},
  {Burgett}, {Chambers}, {Grav}, {Heasley}, {Kaiser}, {Kudritzki}, {Magnier},
  {Morgan}, \& {Price}}]{Gezari12}
{Gezari}, S., {Chornock}, R., {Rest}, A., {et~al.} 2012,
  \href{http://dx.doi.org/10.1038/nature10990}{\color{magenta}\nat},
  \href{http://adsabs.harvard.edu/abs/2012Natur.485..217G}{\color{blue}485},
  \href{http://adsabs.harvard.edu/abs/2012Natur.485..217G}{\color{blue}217}

\bibitem[{{Goldstein} {et~al.}(2015){Goldstein}, {D'Andrea}, {Fischer},
  {Foley}, {Gupta}, {Kessler}, {Kim}, {Nichol}, {Nugent}, {Papadopoulos},
  {Sako}, {Smith}, {Sullivan}, {Thomas}, {Wester}, {Wolf}, {Abdalla},
  {Banerji}, {Benoit-L{\'e}vy}, {Bertin}, {Brooks}, {Carnero Rosell},
  {Castander}, {da Costa}, {Covarrubias}, {DePoy}, {Desai}, {Diehl}, {Doel},
  {Eifler}, {Fausti Neto}, {Finley}, {Flaugher}, {Fosalba}, {Frieman},
  {Gerdes}, {Gruen}, {Gruendl}, {James}, {Kuehn}, {Kuropatkin}, {Lahav}, {Li},
  {Maia}, {Makler}, {March}, {Marshall}, {Martini}, {Merritt}, {Miquel},
  {Nord}, {Ogando}, {Plazas}, {Romer}, {Roodman}, {Sanchez}, {Scarpine},
  {Schubnell}, {Sevilla-Noarbe}, {Smith}, {Soares-Santos}, {Sobreira},
  {Suchyta}, {Swanson}, {Tarle}, {Thaler}, \& {Walker}}]{Goldstein15}
{Goldstein}, D.~A., {D'Andrea}, C.~B., {Fischer}, J.~A., {et~al.} 2015,
  \href{http://dx.doi.org/10.1088/0004-6256/150/3/82}{\color{magenta}\aj},
  \href{http://adsabs.harvard.edu/abs/2015AJ....150...82G}{\color{blue}150},
  \href{http://adsabs.harvard.edu/abs/2015AJ....150...82G}{\color{blue}82}

\bibitem[{{Henrion} {et~al.}(2011){Henrion}, {Mortlock}, {Hand}, \&
  {Gandy}}]{Henrion11}
{Henrion}, M., {Mortlock}, D.~J., {Hand}, D.~J., \& {Gandy}, A. 2011,
  \href{http://dx.doi.org/10.1111/j.1365-2966.2010.18055.x}{\color{magenta}\mnras},
  \href{https://ui.adsabs.harvard.edu/abs/2011MNRAS.412.2286H}{\color{blue}412},
  \href{https://ui.adsabs.harvard.edu/abs/2011MNRAS.412.2286H}{\color{blue}2286}

\bibitem[{Hunter(2007)}]{Hunter07}
Hunter, J.~D. 2007,
  \href{http://dx.doi.org/10.1109/MCSE.2007.55}{\color{magenta}Computing In
  Science \& Engineering}, 9, 90

\bibitem[{{IceCube Collaboration} {et~al.}(2018){IceCube Collaboration},
  {Aartsen}, {Ackermann}, {Adams}, {Aguilar}, {Ahlers}, {Ahrens}, {Al Samarai},
  {Altmann}, {Andeen}, {Anderson}, {Ansseau}, {Anton}, {Arg{\"u}elles},
  {Auffenberg}, {Axani}, {Bagherpour}, {Bai}, {Barron}, {Barwick}, {Baum},
  {Bay}, {Beatty}, {Becker Tjus}, {Becker}, {BenZvi}, {Berley}, {Bernardini},
  {Besson}, {Binder}, {Bindig}, {Blaufuss}, {Blot}, {Bohm}, {B{\"o}rner},
  {Bos}, {B{\"o}ser}, {Botner}, {Bourbeau}, {Bourbeau}, {Bradascio}, {Braun},
  {Brenzke}, {Bretz}, {Bron}, {Brostean-Kaiser}, {Burgman}, {Busse}, {Carver},
  {Cheung}, {Chirkin}, {Christov}, {Clark}, {Classen}, {Coenders}, {Collin},
  {Conrad}, {Coppin}, {Correa}, {Cowen}, {Cross}, {Dave}, {Day}, {de
  Andr{\'e}}, {De Clercq}, {DeLaunay}, {Dembinski}, {De Ridder}, {Desiati}, {de
  Vries}, {de Wasseige}, {de With}, {DeYoung}, {D{\'\i}az-V{\'e}lez}, {di
  Lorenzo}, {Dujmovic}, {Dumm}, {Dunkman}, {Dvorak}, {Eberhardt}, {Ehrhardt},
  {Eichmann}, {Eller}, {Evenson}, {Fahey}, {Fazely}, {Felde}, {Filimonov},
  {Finley}, {Flis}, {Franckowiak}, {Friedman}, {Fritz}, {Gaisser}, {Gallagher},
  {Gerhardt}, {Ghorbani}, {Glauch}, {Gl{\"u}senkamp}, {Goldschmidt},
  {Gonzalez}, {Grant}, {Griffith}, {Haack}, {Hallgren}, {Halzen}, {Hanson},
  {Hebecker}, {Heereman}, {Helbing}, {Hellauer}, {Hickford}, {Hignight},
  {Hill}, {Hoffman}, {Hoffmann}, {Hoinka}, {Hokanson-Fasig}, {Hoshina},
  {Huang}, {Huber}, {Hultqvist}, {H{\"u}nnefeld}, {Hussain}, {In}, {Iovine},
  {Ishihara}, {Jacobi}, {Japaridze}, {Jeong}, {Jero}, {Jones}, {Kalaczynski},
  {Kang}, {Kappes}, {Kappesser}, {Karg}, {Karle}, {Katz}, {Kauer}, {Keivani},
  {Kelley}, {Kheirandish}, {Kim}, {Kim}, {Kintscher}, {Kiryluk}, {Kittler},
  {Klein}, {Koirala}, {Kolanoski}, {K{\"o}pke}, {Kopper}, {Kopper},
  {Koschinsky}, {Koskinen}, {Kowalski}, {Krings}, {Kroll}, {Kr{\"u}ckl},
  {Kunwar}, {Kurahashi}, {Kuwabara}, {Kyriacou}, {Labare}, {Lanfranchi},
  {Larson}, {Lauber}, {Leonard}, {Lesiak-Bzdak}, {Leuermann}, {Liu}, {Lozano
  Mariscal}, {Lu}, {L{\"u}nemann}, {Luszczak}, {Madsen}, {Maggi}, {Mahn},
  {Mancina}, {Maruyama}, {Mase}, {Maunu}, {Meagher}, {Medici}, {Meier},
  {Menne}, {Merino}, {Meures}, {Miarecki}, {Micallef}, {Moment{\'e}},
  {Montaruli}, {Moore}, {S}, {Morse}, {Moulai}, {Nahnhauer}, {Nakarmi},
  {Naumann}, {Neer}, {Niederhausen}, {Nowicki}, {Nygren}, {Obertacke Pollmann},
  {Olivas}, {O'Murchadha}, {O'Sullivan}, {Palczewski}, {Pandya}, {Pankova},
  {Peiffer}, {Pepper}, {P{\'e}rez de los Heros}, {Pieloth}, {Pinat}, {Plum},
  {Price}, {Przybylski}, {Raab}, {R{\"a}del}, {Rameez}, {Rauch}, {Rawlins},
  {Rea}, {Reimann}, {Relethford}, {Relich}, {Resconi}, {Rhode}, {Richman},
  {Robertson}, {Rongen}, {Rott}, {Ruhe}, {Ryckbosch}, {Rysewyk}, {Safa},
  {S{\"a}lzer}, {Sanchez Herrera}, {Sandrock}, {Sandroos}, {Santander},
  {Sarkar}, {Sarkar}, {Satalecka}, {Schlunder}, {Schmidt}, {Schneider},
  {Schoenen}, {Sch{\"o}neberg}, {Schumacher}, {Sclafani}, {Seckel},
  {Seunarine}, {Soedingrekso}, {Soldin}, {Song}, {Spiczak}, {Spiering},
  {Stachurska}, {Stamatikos}, {Stanev}, {Stasik}, {Stein}, {Stettner},
  {Steuer}, {Stezelberger}, {Stokstad}, {St{\"o}{\ss}l}, {Strotjohann},
  {Stuttard}, {Sullivan}, {Sutherland}, {Taboada}, {Tatar}, {Tenholt},
  {Ter-Antonyan}, {Terliuk}, {Tilav}, {Toale}, {Tobin}, {Toennis}, {Toscano},
  {Tosi}, {Tselengidou}, {Tung}, {Turcati}, {Turley}, {Ty}, {Unger}, {Usner},
  {Vandenbroucke}, {Van Driessche}, {van Eijk}, {van Eijndhoven}, {Vanheule},
  {van Santen}, {Vogel}, {Vraeghe}, {Walck}, {Wallace}, {Wallraff}, {Wandler},
  {Wandkowsky}, {Waza}, {Weaver}, {Weiss}, {Wendt}, {Werthebach}, {Westerhoff},
  {Whelan}, {Whitehorn}, {Wiebe}, {Wiebusch}, {Wille}, {Williams}, {Wills},
  {Wolf}, {Wood}, {Wood}, {Woschnagg}, {Xu}, {Xu}, {Xu}, {Yanez}, {Yodh},
  {Yoshida}, {Yuan}, {Fermi-LAT Collaboration}, {Abdollahi}, {Ajello},
  {Angioni}, {Baldini}, {Ballet}, {Barbiellini}, {Bastieri}, {Bechtol},
  {Bellazzini}, {Berenji}, {Bissaldi}, {Blandford}, {Bonino}, {Bottacini},
  {Bregeon}, {Bruel}, {Buehler}, {Burnett}, {Burns}, {Buson}, {Cameron},
  {Caputo}, {Caraveo}, {Cavazzuti}, {Charles}, {Chen}, {Cheung}, {Chiang},
  {Chiaro}, {Ciprini}, {Cohen-Tanugi}, {Conrad}, {Costantin}, {Cutini},
  {D'Ammando}, {de Palma}, {Digel}, {Di Lalla}, {Di Mauro}, {Di Venere},
  {Dom{\'\i}nguez}, {Favuzzi}, {Franckowiak}, {Fukazawa}, {Funk}, {Fusco},
  {Gargano}, {Gasparrini}, {Giglietto}, {Giomi}, {Giommi}, {Giordano},
  {Giroletti}, {Glanzman}, {Green}, {Grenier}, {Grondin}, {Guiriec}, {Harding},
  {Hayashida}, {Hays}, {Hewitt}, {Horan}, {J{\'o}hannesson}, {Kadler},
  {Kensei}, {Kocevski}, {Krauss}, {Kreter}, {Kuss}, {La Mura}, {Larsson},
  {Latronico}, {Lemoine-Goumard}, {Li}, {Longo}, {Loparco}, {Lovellette},
  {Lubrano}, {Magill}, {Maldera}, {Malyshev}, {Manfreda}, {Mazziotta},
  {McEnery}, {Meyer}, {Michelson}, {Mizuno}, {Monzani}, {Morselli},
  {Moskalenko}, {Negro}, {Nuss}, {Ojha}, {Omodei}, {Orienti}, {Orlando},
  {Palatiello}, {Paliya}, {Perkins}, {Persic}, {Pesce-Rollins}, {Piron},
  {Porter}, {Principe}, {Rain{\`o}}, {Rando}, {Rani}, {Razzano}, {Razzaque},
  {Reimer}, {Reimer}, {Renault-Tinacci}, {Ritz}, {Rochester}, {Saz Parkinson},
  {Sgr{\`o}}, {Siskind}, {Spandre}, {Spinelli}, {Suson}, {Tajima}, {Takahashi},
  {Tanaka}, {Thayer}, {Thompson}, {Tibaldo}, {Torres}, {Torresi}, {Tosti},
  {Troja}, {Valverde}, {Vianello}, {Vogel}, {Wood}, {Wood}, {Zaharijas}, {MAGIC
  Collaboration}, {Ahnen}, {Ansoldi}, {Antonelli}, {Arcaro}, {Baack},
  {Babi{\'c}}, {Banerjee}, {Bangale}, {Barres de Almeida}, {Barrio}, {Becerra
  Gonz{\'a}lez}, {Bednarek}, {Bernardini}, {Berti}, {Bhattacharyya}, {Biland},
  {Blanch}, {Bonnoli}, {Carosi}, {Carosi}, {Ceribella}, {Chatterjee}, {Colak},
  {Colin}, {Colombo}, {Contreras}, {Cortina}, {Covino}, {Cumani}, {Da Vela},
  {Dazzi}, {De Angelis}, {De Lotto}, {Delfino}, {Delgado}, {Di Pierro},
  {Dom{\'\i}nguez}, {Dominis Prester}, {Dorner}, {Doro}, {Einecke},
  {Elsaesser}, {Fallah Ramazani}, {Fern{\'a}ndez-Barral}, {Fidalgo}, {Foffano},
  {Pfrang}, {Fonseca}, {Font}, {Franceschini}, {Fruck}, {Galindo}, {Gallozzi},
  {Garc{\'\i}a L{\'o}pez}, {Garczarczyk}, {Gaug}, {Giammaria}, {Godinovi{\'c}},
  {Gora}, {Guberman}, {Hadasch}, {Hahn}, {Hassan}, {Hayashida}, {Herrera},
  {Hose}, {Hrupec}, {Inoue}, {Ishio}, {Konno}, {Kubo}, {Kushida}, {Lelas},
  {Lindfors}, {Lombardi}, {Longo}, {L{\'o}pez}, {Maggio}, {Majumdar},
  {Makariev}, {Maneva}, {Manganaro}, {Mannheim}, {Maraschi}, {Mariotti},
  {Mart{\'\i}nez}, {Masuda}, {Mazin}, {Minev}, {M}, {Mirzoyan}, {Moralejo},
  {Moreno}, {Moretti}, {Nagayoshi}, {Neustroev}, {Niedzwiecki}, {Nievas
  Rosillo}, {Nigro}, {Nilsson}, {Ninci}, {Nishijima}, {Noda}, {Nogu{\'e}s},
  {Paiano}, {Palacio}, {Paneque}, {Paoletti}, {Paredes}, {Pedaletti},
  {Peresano}, {Persic}, {Prada Moroni}, {Prandini}, {Puljak}, {Rodriguez
  Garcia}, {Reichardt}, {Rhode}, {Rib{\'o}}, {Rico}, {Righi}, {Rugliancich},
  {Saito}, {Satalecka}, {Schweizer}, {Sitarek}, {{\v{S}}nidaric ́},
  {Sobczynska}, {Stamerra}, {Strzys}, {Suri{\'c}}, {Takahashi}, {Tavecchio},
  {Temnikov}, {Terzi{\'c}}, {Teshima}, {Torres-Alb{\`a}}, {Treves},
  {Tsujimoto}, {Vanzo}, {Vazquez Acosta}, {Vovk}, {Ward}, {Will}, {S}, {Zaric
  ́}, {AGILE Team}, {Lucarelli}, {Tavani}, {Piano}, {Donnarumma}, {Pittori},
  {Verrecchia}, {Barbiellini}, {Bulgarelli}, {Caraveo}, {Cattaneo},
  {Colafrancesco}, {Costa}, {Di Cocco}, {Ferrari}, {Gianotti}, {Giuliani},
  {Lipari}, {Mereghetti}, {Morselli}, {Pacciani}, {Paoletti}, {Parmiggiani},
  {Pellizzoni}, {Picozza}, {Pilia}, {Rappoldi}, {Trois}, {Vercellone},
  {Vittorini}, {ASAS-SN Team}, {Stanek}, {Kochanek}, {Beacom}, {Thompson},
  {Holoien}, {Dong}, {Prieto}, {Shappee}, {Holmbo}, {HAWC Collaboration},
  {Abeysekara}, {Albert}, {Alfaro}, {Alvarez}, {Arceo},
  {Arteaga-Vel{\'a}zquez}, {Avila Rojas}, {Ayala Solares}, {Becerril},
  {Belmont-Moreno}, {Bernal}, {Caballero-Mora}, {Capistr{\'a}n},
  {Carrami{\~n}ana}, {Casanova}, {Castillo}, {Cotti}, {Cotzomi}, {Couti{\~n}o
  de Le{\'o}n}, {De Le{\'o}n}, {De la Fuente}, {Diaz Hernandez}, {Dichiara},
  {Dingus}, {DuVernois}, {D{\'\i}az-V{\'e}lez}, {Ellsworth}, {Engel},
  {Fiorino}, {Fleischhack}, {Fraija}, {Garc{\'\i}a-Gonz{\'a}lez}, {Garfias},
  {Gonz{\'a}lez Mu{\~n}oz}, {Gonz{\'a}lez}, {Goodman}, {Hampel-Arias},
  {Harding}, {Hernand ez}, {Hona}, {Hueyotl-Zahuantitla}, {Hui},
  {H{\"u}ntemeyer}, {Iriarte}, {Jardin-Blicq}, {Joshi}, {Kaufmann}, {Kunde},
  {Lara}, {Lauer}, {Lee}, {Lennarz}, {Le{\'o}n Vargas}, {Linnemann},
  {Longinotti}, {Luis-Raya}, {Luna-Garc{\'\i}a}, {Malone}, {Marinelli},
  {Martinez}, {Martinez-Castellanos}, {Mart{\'\i}nez-Castro},
  {Mart{\'\i}nez-Huerta}, {Matthews}, {Miranda-Romagnoli}, {Moreno},
  {Mostaf{\'a}}, {Nayerhoda}, {Nellen}, {Newbold}, {Nisa}, {Noriega-Papaqui},
  {Pelayo}, {Pretz}, {P{\'e}rez-P{\'e}rez}, {Ren}, {Rho}, {Rivi{\`e}re},
  {Rosa-Gonz{\'a}lez}, {Rosenberg}, {Ruiz-Velasco}, {Ruiz-Velasco}, {Salesa
  Greus}, {Sandoval}, {Schneider}, {Schoorlemmer}, {Sinnis}, {Smith},
  {Springer}, {Surajbali}, {Tibolla}, {Tollefson}, {Torres}, {Villase{\~n}or},
  {Weisgarber}, {Werner}, {Yapici}, {Gaurang}, {Zepeda}, {Zhou}, {{\'A}lvarez},
  {H.~E.~S.~S. Collaboration}, {Abdalla}, {Ang{\"u}ner}, {Armand}, {Backes},
  {Becherini}, {Berge}, {B{\"o}ttcher}, {Boisson}, {Bolmont}, {Bonnefoy},
  {Bordas}, {Brun}, {B{\"u}chele}, {Bulik}, {Caroff}, {Carosi}, {Casanova},
  {Cerruti}, {Chakraborty}, {Chandra}, {Chen}, {Colafrancesco}, {Davids},
  {Deil}, {Devin}, {Djannati-Ata{\"\i}}, {Egberts}, {Emery}, {Eschbach},
  {Fiasson}, {Fontaine}, {Funk}, {F{\"u}{\ss}ling}, {Gallant}, {Gat{\'e}},
  {Giavitto}, {Glawion}, {Glicenstein}, {Gottschall}, {Grondin}, {Haupt},
  {Henri}, {Hinton}, {Hoischen}, {Holch}, {Huber}, {Jamrozy}, {Jankowsky},
  {Jankowsky}, {Jouvin}, {Jung-Richardt}, {Kerszberg}, {Kh{\'e}lifi}, {King},
  {Klepser}, {Kluz ́niak}, {Komin}, {Kraus}, {Lefaucheur}, {Lemi{\`e}re},
  {Lemoine-Goumard}, {Lenain}, {Leser}, {Lohse}, {L{\'o}pez-Coto}, {Lorentz},
  {Lypova}, {Marandon}, {Guillem Mart{\'\i}-Devesa}, {Maurin}, {Mitchell},
  {Moderski}, {Mohamed}, {Mohrmann}, {Moulin}, {Murach}, {de Naurois},
  {Niederwanger}, {Niemiec}, {Oakes}, {O'Brien}, {Ohm}, {Ostrowski}, {Oya},
  {Panter}, {Parsons}, {Perennes}, {Piel}, {Pita}, {Poireau}, {Priyana Noel},
  {Prokoph}, {P{\"u}hlhofer}, {Quirrenbach}, {Raab}, {Rauth}, {Renaud},
  {Rieger}, {Rinchiuso}, {Romoli}, {Rowell}, {Rudak}, {Sasaki}, {Sanchez},
  {Schlickeiser}, {Sch{\"u}ssler}, {Schulz}, {Schwanke}, {Seglar-Arroyo},
  {Shafi}, {Simoni}, {Sol}, {Stegmann}, {Steppa}, {Tavernier}, {Taylor},
  {Tiziani}, {Trichard}, {Tsirou}, {van Eldik}, {van Rensburg}, {van Soelen},
  {Veh}, {Vincent}, {Voisin}, {Wagner}, {Wagner}, {Wierzcholska}, {Zanin},
  {Zdziarski}, {Zech}, {Ziegler}, {Zorn}, {{\.Z}ywucka}, {INTEGRAL Team},
  {Savchenko}, {Ferrigno}, {Bazzano}, {Diehl}, {Kuulkers}, {Laurent},
  {Mereghetti}, {Natalucci}, {Panessa}, {Rodi}, {Ubertini}, {Kanata}, Teams,
  {Morokuma}, {Ohta}, {Tanaka}, {Mori}, {Yamanaka}, {Kawabata}, {Utsumi},
  {Nakaoka}, {Kawabata}, {Nagashima}, {Yoshida}, {Matsuoka}, {Itoh}, {Kapteyn
  Team}, {Keel}, {Liverpool Telescope Team}, {Copperwheat}, {Steele},
  {Swift/NuSTAR Team}, {Cenko}, {Cowen}, {DeLaunay}, {Evans}, {Fox}, {Keivani},
  {Kennea}, {Marshall}, {Osborne}, {Santander}, {Tohuvavohu}, {Turley},
  {VERITAS Collaboration}, {Abeysekara}, {Archer}, {Benbow}, {Bird}, {Brill},
  {Brose}, {Buchovecky}, {Buckley}, {Bugaev}, {Christiansen}, {Connolly},
  {Cui}, {Daniel}, {Errando}, {Falcone}, {Feng}, {Finley}, {Fortson},
  {Furniss}, {Gueta}, {H{\"u}tten}, {Hervet}, {Hughes}, {Humensky}, {Johnson},
  {Kaaret}, {Kar}, {Kelley-Hoskins}, {Kertzman}, {Kieda}, {Krause},
  {Krennrich}, {Kumar}, {Lang}, {Lin}, {Maier}, {McArthur}, {Moriarty},
  {Mukherjee}, {Nieto}, {O'Brien}, {Ong}, {Otte}, {Park}, {Petrashyk}, {Pohl},
  {Popkow}, {Pueschel}, {Quinn}, {Ragan}, {Reynolds}, {Richards}, {Roache},
  {Rulten}, {Sadeh}, {Santander}, {Scott}, {Sembroski}, {Shahinyan}, {Sushch},
  {Tr{\'e}panier}, {Tyler}, {Vassiliev}, {Wakely}, {Weinstein}, {Wells},
  {Wilcox}, {Wilhelm}, {Williams}, {Zitzer}, {VLA/B Team}, {Tetarenko},
  {Kimball}, {Miller-Jones}, \& {Sivakoff}}]{IceCube-Collaboration18}
{IceCube Collaboration}, {Aartsen}, M.~G., {Ackermann}, M., {et~al.} 2018,
  \href{http://dx.doi.org/10.1126/science.aat1378}{\color{magenta}Science},
  \href{https://ui.adsabs.harvard.edu/abs/2018Sci...361.1378I}{\color{blue}361},
  \href{https://ui.adsabs.harvard.edu/abs/2018Sci...361.1378I}{\color{blue}eaat1378}

\bibitem[{{Ivezi{\'c}} {et~al.}(2019){Ivezi{\'c}}, {Kahn}, {Tyson}, {Abel},
  {Acosta}, {Allsman}, {Alonso}, {AlSayyad}, {Anderson}, {Andrew}, {Angel},
  {Angeli}, {Ansari}, {Antilogus}, {Araujo}, {Armstrong}, {Arndt}, {Astier},
  {Aubourg}, {Auza}, {Axelrod}, {Bard}, {Barr}, {Barrau}, {Bartlett}, {Bauer},
  {Bauman}, {Baumont}, {Bechtol}, {Bechtol}, {Becker}, {Becla}, {Beldica},
  {Bellavia}, {Bianco}, {Biswas}, {Blanc}, {Blazek}, {Bland ford}, {Bloom},
  {Bogart}, {Bond}, {Booth}, {Borgland}, {Borne}, {Bosch}, {Boutigny},
  {Brackett}, {Bradshaw}, {Brand t}, {Brown}, {Bullock}, {Burchat}, {Burke},
  {Cagnoli}, {Calabrese}, {Callahan}, {Callen}, {Carlin}, {Carlson}, {Chand
  rasekharan}, {Charles-Emerson}, {Chesley}, {Cheu}, {Chiang}, {Chiang},
  {Chirino}, {Chow}, {Ciardi}, {Claver}, {Cohen-Tanugi}, {Cockrum}, {Coles},
  {Connolly}, {Cook}, {Cooray}, {Covey}, {Cribbs}, {Cui}, {Cutri}, {Daly},
  {Daniel}, {Daruich}, {Daubard}, {Daues}, {Dawson}, {Delgado}, {Dellapenna},
  {de Peyster}, {de Val-Borro}, {Digel}, {Doherty}, {Dubois},
  {Dubois-Felsmann}, {Durech}, {Economou}, {Eifler}, {Eracleous}, {Emmons},
  {Fausti Neto}, {Ferguson}, {Figueroa}, {Fisher-Levine}, {Focke}, {Foss},
  {Frank}, {Freemon}, {Gangler}, {Gawiser}, {Geary}, {Gee}, {Geha}, {Gessner},
  {Gibson}, {Gilmore}, {Glanzman}, {Glick}, {Goldina}, {Goldstein}, {Goodenow},
  {Graham}, {Gressler}, {Gris}, {Guy}, {Guyonnet}, {Haller}, {Harris},
  {Hascall}, {Haupt}, {Hernand ez}, {Herrmann}, {Hileman}, {Hoblitt},
  {Hodgson}, {Hogan}, {Howard}, {Huang}, {Huffer}, {Ingraham}, {Innes},
  {Jacoby}, {Jain}, {Jammes}, {Jee}, {Jenness}, {Jernigan}, {Jevremovi{\'c}},
  {Johns}, {Johnson}, {Johnson}, {Jones}, {Juramy-Gilles}, {Juri{\'c}},
  {Kalirai}, {Kallivayalil}, {Kalmbach}, {Kantor}, {Karst}, {Kasliwal},
  {Kelly}, {Kessler}, {Kinnison}, {Kirkby}, {Knox}, {Kotov}, {Krabbendam},
  {Krughoff}, {Kub{\'a}nek}, {Kuczewski}, {Kulkarni}, {Ku}, {Kurita}, {Lage},
  {Lambert}, {Lange}, {Langton}, {Le Guillou}, {Levine}, {Liang}, {Lim},
  {Lintott}, {Long}, {Lopez}, {Lotz}, {Lupton}, {Lust}, {MacArthur}, {Mahabal},
  {Mand elbaum}, {Markiewicz}, {Marsh}, {Marshall}, {Marshall}, {May},
  {McKercher}, {McQueen}, {Meyers}, {Migliore}, {Miller}, {Mills}, {Miraval},
  {Moeyens}, {Moolekamp}, {Monet}, {Moniez}, {Monkewitz}, {Montgomery},
  {Morrison}, {Mueller}, {Muller}, {Mu{\~n}oz Arancibia}, {Neill}, {Newbry},
  {Nief}, {Nomerotski}, {Nordby}, {O'Connor}, {Oliver}, {Olivier}, {Olsen},
  {O'Mullane}, {Ortiz}, {Osier}, {Owen}, {Pain}, {Palecek}, {Parejko},
  {Parsons}, {Pease}, {Peterson}, {Peterson}, {Petravick}, {Libby Petrick},
  {Petry}, {Pierfederici}, {Pietrowicz}, {Pike}, {Pinto}, {Plante}, {Plate},
  {Plutchak}, {Price}, {Prouza}, {Radeka}, {Rajagopal}, {Rasmussen},
  {Regnault}, {Reil}, {Reiss}, {Reuter}, {Ridgway}, {Riot}, {Ritz}, {Robinson},
  {Roby}, {Roodman}, {Rosing}, {Roucelle}, {Rumore}, {Russo}, {Saha},
  {Sassolas}, {Schalk}, {Schellart}, {Schindler}, {Schmidt}, {Schneider},
  {Schneider}, {Schoening}, {Schumacher}, {Schwamb}, {Sebag}, {Selvy},
  {Sembroski}, {Seppala}, {Serio}, {Serrano}, {Shaw}, {Shipsey}, {Sick},
  {Silvestri}, {Slater}, {Smith}, {Smith}, {Sobhani}, {Soldahl},
  {Storrie-Lombardi}, {Stover}, {Strauss}, {Street}, {Stubbs}, {Sullivan},
  {Sweeney}, {Swinbank}, {Szalay}, {Takacs}, {Tether}, {Thaler}, {Thayer},
  {Thomas}, {Thornton}, {Thukral}, {Tice}, {Trilling}, {Turri}, {Van Berg},
  {Vanden Berk}, {Vetter}, {Virieux}, {Vucina}, {Wahl}, {Walkowicz}, {Walsh},
  {Walter}, {Wang}, {Wang}, {Warner}, {Wiecha}, {Willman}, {Winters},
  {Wittman}, {Wolff}, {Wood-Vasey}, {Wu}, {Xin}, {Yoachim}, \&
  {Zhan}}]{Ivezic19}
{Ivezi{\'c}}, {\v{Z}}., {Kahn}, S.~M., {Tyson}, J.~A., {et~al.} 2019,
  \href{http://dx.doi.org/10.3847/1538-4357/ab042c}{\color{magenta}\apj},
  \href{https://ui.adsabs.harvard.edu/abs/2019ApJ...873..111I}{\color{blue}873},
  \href{https://ui.adsabs.harvard.edu/abs/2019ApJ...873..111I}{\color{blue}111}

\bibitem[{{Jones} {et~al.}(2020){Jones}, {Foley}, {Narayan}, {Hjorth}, {Huber},
  {Aleo}, {Alexander}, {Angus}, {Auchettl}, {Baldassare}, {Bruun}, {Chambers},
  {Chatterjee}, {Coppejans}, {Coulter}, {DeMarchi}, {Dimitriadis}, {Drout},
  {Engel}, {French}, {Gagliano}, {Gall}, {Hung}, {Izzo}, {Jacobson-Gal{\'a}n},
  {Kilpatrick}, {Korhonen}, {Margutti}, {Raimundo}, {Ramirez-Ruiz}, {Rest},
  {Rojas-Bravo}, {Schultz}, {Siebert}, {Smartt}, {Smith}, {Terreran}, {Wang},
  {Wojtak}, {Agnello}, {Ansari}, {Arendse}, {Baldeschi}, {Blanchard},
  {Brethauer}, {Bright}, {Brown}, {deBoer}, {Dodd}, {Fairlamb}, {Grillo},
  {Hajela}, {Hede}, {Kolborg}, {Law-Smith}, {Lin}, {Magnier}, {Malanchev},
  {Matthews}, {Mockler}, {Muthukrishna}, {Pan}, {Pfister}, {Ramanah}, {Rest},
  {Sarangi}, {Schr{\o}der}, {Stauffer}, {Stroh}, {Taggart}, {Tinyanont}, \&
  {Wainscoat}}]{Jones20}
{Jones}, D.~O., {Foley}, R.~J., {Narayan}, G., {et~al.} 2020,
  \href{https://arxiv.org/abs/2010.09724}{\color{magenta}arXiv},
  \href{https://ui.adsabs.harvard.edu/abs/2020arXiv201009724J}{\color{blue}arXiv:2010.09724}

\bibitem[{{Kaiser} {et~al.}(1995){Kaiser}, {Squires}, \&
  {Broadhurst}}]{Kaiser95}
{Kaiser}, N., {Squires}, G., \& {Broadhurst}, T. 1995,
  \href{http://dx.doi.org/10.1086/176071}{\color{magenta}\apj},
  \href{https://ui.adsabs.harvard.edu/abs/1995ApJ...449..460K}{\color{blue}449},
  \href{https://ui.adsabs.harvard.edu/abs/1995ApJ...449..460K}{\color{blue}460}

\bibitem[{{Kulkarni}(2020)}]{Kulkarni20}
{Kulkarni}, S.~R. 2020,
  \href{https://arxiv.org/abs/2004.03511}{\color{magenta}arXiv},
  \href{https://ui.adsabs.harvard.edu/abs/2020arXiv200403511K}{\color{blue}arXiv:2004.03511}

\bibitem[{{Leauthaud} {et~al.}(2007){Leauthaud}, {Massey}, {Kneib}, {Rhodes},
  {Johnston}, {Capak}, {Heymans}, {Ellis}, {Koekemoer}, {Le F{\`e}vre},
  {Mellier}, {R{\'e}fr{\'e}gier}, {Robin}, {Scoville}, {Tasca}, {Taylor}, \&
  {Van Waerbeke}}]{Leauthaud07}
{Leauthaud}, A., {Massey}, R., {Kneib}, J.-P., {et~al.} 2007,
  \href{http://dx.doi.org/10.1086/516598}{\color{magenta}\apjs},
  \href{http://adsabs.harvard.edu/abs/2007ApJS..172..219L}{\color{blue}172},
  \href{http://adsabs.harvard.edu/abs/2007ApJS..172..219L}{\color{blue}219}

\bibitem[{{Lupton} {et~al.}(2001){Lupton}, {Gunn}, {Ivezi{\'c}}, {Knapp}, \&
  {Kent}}]{Lupton01}
{Lupton}, R., {Gunn}, J.~E., {Ivezi{\'c}}, Z., {et~al.} 2001, Astronomical
  Society of the Pacific Conference Series,
  \href{http://adsabs.harvard.edu/abs/2001ASPC..238..269L}{\color{blue}238},
  \href{http://adsabs.harvard.edu/abs/2001ASPC..238..269L}{\color{blue}269}

\bibitem[{{Masci} {et~al.}(2017){Masci}, {Laher}, {Rebbapragada}, {Doran},
  {Miller}, {Bellm}, {Kasliwal}, {Ofek}, {Surace}, {Shupe}, {Grillmair},
  {Jackson}, {Barlow}, {Yan}, {Cao}, {Cenko}, {Storrie-Lombardi}, {Helou},
  {Prince}, \& {Kulkarni}}]{Masci17}
{Masci}, F.~J., {Laher}, R.~R., {Rebbapragada}, U.~D., {et~al.} 2017,
  \href{http://dx.doi.org/10.1088/1538-3873/129/971/014002}{\color{magenta}\pasp},
  \href{http://adsabs.harvard.edu/abs/2017PASP..129a4002M}{\color{blue}129},
  \href{http://adsabs.harvard.edu/abs/2017PASP..129a4002M}{\color{blue}014002}

\bibitem[{{Masci} {et~al.}(2019){Masci}, {Laher}, {Rusholme}, {Shupe}, {Groom},
  {Surace}, {Jackson}, {Monkewitz}, {Beck}, {Flynn}, {Terek}, {Landry},
  {Hacopians}, {Desai}, {Howell}, {Brooke}, {Imel}, {Wachter}, {Ye}, {Lin},
  {Cenko}, {Cunningham}, {Rebbapragada}, {Bue}, {Miller}, {Mahabal}, {Bellm},
  {Patterson}, {Juri{\'c}}, {Golkhou}, {Ofek}, {Walters}, {Graham}, {Kasliwal},
  {Dekany}, {Kupfer}, {Burdge}, {Cannella}, {Barlow}, {Van Sistine}, {Giomi},
  {Fremling}, {Blagorodnova}, {Levitan}, {Riddle}, {Smith}, {Helou}, {Prince},
  \& {Kulkarni}}]{Masci19}
{Masci}, F.~J., {Laher}, R.~R., {Rusholme}, B., {et~al.} 2019,
  \href{http://dx.doi.org/10.1088/1538-3873/aae8ac}{\color{magenta}\pasp},
  \href{https://ui.adsabs.harvard.edu/abs/2019PASP..131a8003M}{\color{blue}131},
  \href{https://ui.adsabs.harvard.edu/abs/2019PASP..131a8003M}{\color{blue}018003}

\bibitem[{McKinney(2010)}]{McKinney10}
McKinney, W. 2010, 56

\bibitem[{{Miller} {et~al.}(2017){Miller}, {Kulkarni}, {Cao}, {Laher}, {Masci},
  \& {Surace}}]{Miller17}
{Miller}, A.~A., {Kulkarni}, M.~K., {Cao}, Y., {et~al.} 2017,
  \href{http://dx.doi.org/10.3847/1538-3881/153/2/73}{\color{magenta}\aj},
  \href{http://adsabs.harvard.edu/abs/2017AJ....153...73M}{\color{blue}153},
  \href{http://adsabs.harvard.edu/abs/2017AJ....153...73M}{\color{blue}73}

\bibitem[{{M{\"o}ller} {et~al.}(2020){M{\"o}ller}, {Peloton}, {Ishida},
  {Arnault}, {Bachelet}, {Blaineau}, {Boutigny}, {Chauhan}, {Gangler},
  {Hernandez}, {Hrivnac}, {Leoni}, {Leroy}, {Moniez}, {Pateyron}, {Ramparison},
  {Turpin}, {Ansari}, {Allam}, {Bajat}, {Biswas}, {Boucaud}, {Bregeon},
  {Campagne}, {Cohen-Tanugi}, {Coleiro}, {Dornic}, {Fouchez}, {Godet}, {Gris},
  {Karpov}, {Nebot Gomez-Moran}, {Neveu}, {Plaszczynski}, {Savchenko}, \&
  {Webb}}]{Moller20}
{M{\"o}ller}, A., {Peloton}, J., {Ishida}, E. E.~O., {et~al.} 2020,
  \href{https://arxiv.org/abs/2009.10185}{\color{magenta}arXiv},
  \href{https://ui.adsabs.harvard.edu/abs/2020arXiv200910185M}{\color{blue}arXiv:2009.10185}

\bibitem[{{Patterson} {et~al.}(2019){Patterson}, {Bellm}, {Rusholme}, {Masci},
  {Juric}, {Krughoff}, {Golkhou}, {Graham}, {Kulkarni}, {Helou}, \& {Zwicky
  Transient Facility Collaboration}}]{Patterson19}
{Patterson}, M.~T., {Bellm}, E.~C., {Rusholme}, B., {et~al.} 2019,
  \href{http://dx.doi.org/10.1088/1538-3873/aae904}{\color{magenta}\pasp},
  \href{https://ui.adsabs.harvard.edu/abs/2019PASP..131a8001P}{\color{blue}131},
  \href{https://ui.adsabs.harvard.edu/abs/2019PASP..131a8001P}{\color{blue}018001}

\bibitem[{Pedregosa {et~al.}(2011)Pedregosa, Varoquaux, Gramfort, Michel,
  Thirion, Grisel, Blondel, Prettenhofer, Weiss, Dubourg, Vanderplas, Passos,
  Cournapeau, Brucher, Perrot, \& Duchesnay}]{Pedregosa11}
Pedregosa, F., Varoquaux, G., Gramfort, A., {et~al.} 2011, Journal of Machine
  Learning Research, 12, 2825

\bibitem[{{Perryman} {et~al.}(2001){Perryman}, {de Boer}, {Gilmore}, {H{\o}g},
  {Lattanzi}, {Lindegren}, {Luri}, {Mignard}, {Pace}, \& {de
  Zeeuw}}]{Perryman01}
{Perryman}, M.~A.~C., {de Boer}, K.~S., {Gilmore}, G., {et~al.} 2001,
  \href{http://dx.doi.org/10.1051/0004-6361:20010085}{\color{magenta}\aap},
  \href{http://adsabs.harvard.edu/abs/2001A%26A...369..339P}{\color{blue}369},
  \href{http://adsabs.harvard.edu/abs/2001A%26A...369..339P}{\color{blue}339}

\bibitem[{{Prentice} {et~al.}(2018){Prentice}, {Maguire}, {Smartt}, {Magee},
  {Schady}, {Sim}, {Chen}, {Clark}, {Colin}, {Fulton}, {McBrien}, {O'Neill},
  {Smith}, {Ashall}, {Chambers}, {Denneau}, {Flewelling}, {Heinze}, {Holoien},
  {Huber}, {Kochanek}, {Mazzali}, {Prieto}, {Rest}, {Shappee}, {Stalder},
  {Stanek}, {Stritzinger}, {Thompson}, \& {Tonry}}]{Prentice18}
{Prentice}, S.~J., {Maguire}, K., {Smartt}, S.~J., {et~al.} 2018,
  \href{http://dx.doi.org/10.3847/2041-8213/aadd90}{\color{magenta}\apjl},
  \href{https://ui.adsabs.harvard.edu/abs/2018ApJ...865L...3P}{\color{blue}865},
  \href{https://ui.adsabs.harvard.edu/abs/2018ApJ...865L...3P}{\color{blue}L3}

\bibitem[{{Quimby} {et~al.}(2011){Quimby}, {Kulkarni}, {Kasliwal}, {Gal-Yam},
  {Arcavi}, {Sullivan}, {Nugent}, {Thomas}, {Howell}, {Nakar}, {Bildsten},
  {Theissen}, {Law}, {Dekany}, {Rahmer}, {Hale}, {Smith}, {Ofek}, {Zolkower},
  {Velur}, {Walters}, {Henning}, {Bui}, {McKenna}, {Poznanski}, {Cenko}, \&
  {Levitan}}]{quimby11}
{Quimby}, R.~M., {Kulkarni}, S.~R., {Kasliwal}, M.~M., {et~al.} 2011,
  \href{http://dx.doi.org/10.1038/nature10095}{\color{magenta}\nat},
  \href{http://adsabs.harvard.edu/abs/2011Natur.474..487Q}{\color{blue}474},
  \href{http://adsabs.harvard.edu/abs/2011Natur.474..487Q}{\color{blue}487}

\bibitem[{{Slater} {et~al.}(2020){Slater}, {Ivezi{\'c}}, \&
  {Lupton}}]{Slater20}
{Slater}, C.~T., {Ivezi{\'c}}, {\v{Z}}., \& {Lupton}, R.~H. 2020,
  \href{http://dx.doi.org/10.3847/1538-3881/ab6166}{\color{magenta}\aj},
  \href{https://ui.adsabs.harvard.edu/abs/2020AJ....159...65S}{\color{blue}159},
  \href{https://ui.adsabs.harvard.edu/abs/2020AJ....159...65S}{\color{blue}65}

\bibitem[{{Smith} {et~al.}(2020){Smith}, {Smartt}, {Young}, {Tonry}, {Denneau},
  {Flewelling}, {Heinze}, {Weiland}, {Stalder}, {Rest}, {Stubbs}, {Anderson},
  {Chen}, {Clark}, {Do}, {F{\"o}rster}, {Fulton}, {Gillanders}, {McBrien},
  {O'Neill}, {Srivastav}, \& {Wright}}]{Smith20}
{Smith}, K.~W., {Smartt}, S.~J., {Young}, D.~R., {et~al.} 2020,
  \href{http://dx.doi.org/10.1088/1538-3873/ab936e}{\color{magenta}\pasp},
  \href{https://ui.adsabs.harvard.edu/abs/2020PASP..132h5002S}{\color{blue}132},
  \href{https://ui.adsabs.harvard.edu/abs/2020PASP..132h5002S}{\color{blue}085002}

\bibitem[{{Tachibana} \& {Miller}(2018)}]{Tachibana18}
{Tachibana}, Y., \& {Miller}, A.~A. 2018,
  \href{http://dx.doi.org/10.1088/1538-3873/aae3d9}{\color{magenta}\pasp},
  \href{http://adsabs.harvard.edu/abs/2018PASP..130l8001T}{\color{blue}130},
  \href{http://adsabs.harvard.edu/abs/2018PASP..130l8001T}{\color{blue}128001}

\bibitem[{{Virtanen} {et~al.}(2020){Virtanen}, {Gommers}, {Oliphant},
  {Haberland}, {Reddy}, {Cournapeau}, {Burovski}, {Peterson}, {Weckesser},
  {Bright}, {van der Walt}, {Brett}, {Wilson}, {Jarrod Millman}, {Mayorov},
  {Nelson}, {Jones}, {Kern}, {Larson}, {Carey}, {Polat}, {Feng}, {Moore}, {Vand
  erPlas}, {Laxalde}, {Perktold}, {Cimrman}, {Henriksen}, {Quintero}, {Harris},
  {Archibald}, {Ribeiro}, {Pedregosa}, {van Mulbregt}, \&
  {Contributors}}]{2020SciPy-NMeth}
{Virtanen}, P., {Gommers}, R., {Oliphant}, T.~E., {et~al.} 2020,
  \href{http://dx.doi.org/https://doi.org/10.1038/s41592-019-0686-2}{\color{magenta}Nature
  Methods}, \href{https://rdcu.be/b08Wh}{\color{blue}17},
  \href{https://rdcu.be/b08Wh}{\color{blue}261}

\end{thebibliography}

%% Include this line if you are using the \added, \replaced, \deleted
%% commands to see a summary list of all changes at the end of the article.
%\listofchanges

\end{document}